\documentclass[12pt]{article}
\usepackage{latexsym}
\usepackage{amsmath}
\usepackage{amssymb}
\usepackage{epsfig,graphics}
\usepackage{graphicx}
\usepackage{booktabs}
\usepackage{multirow}
\usepackage{caption}
\usepackage{subcaption}
\usepackage{tikz}

\newcommand{\be}{\begin{equation}}
\newcommand{\ee}{\end{equation}}

\numberwithin{equation}{section}

\begin{document}

\begin{titlepage}

\vspace*{0.6in}

\begin{center}
  {\large\bf More methods for calculating the topological charge (density) \\
    of SU(N) lattice gauge fields in 3+1 dimensions} \\
\vspace*{0.75in}
{Michael Teper\\
\vspace*{.25in}
Rudolf Peierls Centre for Theoretical Physics, University of Oxford,\\
Parks Road, Oxford OX1 3PU, UK\\
\centerline{and}
All Souls College, University of Oxford,\\
High Street, Oxford OX1 4AL, UK}
\end{center}

\vspace*{0.6in}

\begin{center}
{\bf Abstract}
\end{center}

We revisit old ideas that smearing or blocking an $SU(N)$ lattice gauge field,
or averaging over an ensemble of fields created in the neighbourhood of that field, can
reduce the high frequency fluctuations sufficiently that the naive lattice operator for
the topological charge density is able to provide a reliable measure of the topological charge of
the field. We show that these three methods do indeed provide additional simple methods for
calculating the total topological charge, with smearing particularly economical at current
couplings. More interestingly, the ensemble average method can also be used to  expose the
distribution in space-time of the topological charge and this conceptually transparent, albeit
computationally expensive, method provides a useful benchmark against which to compare
other methods. Using this benchmark we find that a few smearing steps are also reliable in exposing 
the distribution in space-time of the topological charge, thus providing a very economical
and simple method for doing so. We also use the same benchmark to determine what is the number
of `cooling' sweeps one needs to perform in order to expose the charge density reliably.

\vspace*{0.95in}

\leftline{{\it E-mail:} mike.teper@physics.ox.ac.uk}

\end{titlepage}

\setcounter{page}{1}
\newpage
\pagestyle{plain}

\tableofcontents

\section{Introduction}
\label{section_intro}

Calculating the topological properties of $SU(N)$ gauge theories in $3+1$ dimensions
is in general a non-perturbative problem and has been an important focus of work
in the lattice gauge theory community since the early days of computer simulations.
There are now several reliable methods for calculating the topological charge of a given
gauge field, although calculating the distribution in space-time of the topological
charge density remains more challenging.
Most of these methods are quite expensive in computational terms and so
there is reason to try and identify additional methods that are economical, and
indeed to see how far one can also calculate the distribution of the topological charge
density in a given gauge field. In this paper we examine some methods that are
closely related to old proposals
\cite{MT-methods}.
The methods which we label by `repetition' and `blocking' appeared promising after some
preliminary trials in semi-realistic contexts. Here we intend to perform large-scale tests
at realistic lattice couplings and volumes. The `blocking' algorithm has also been used
to improve geometric algorithms for calculating the topological charge
\cite{DPMT89}.
The third method, `smearing', is a close cousin of `blocking' and has also been employed
in the past
\cite{AH97,AH98}
albeit in a way that is different to the way we shall use it in this paper.

In the next section we briefly outline our (entirely standard) lattice setup. We
follow this, in Section~\ref{section_topology}, with a brief reminder of some currently
used methods for calculating topology on the lattice. In Section~\ref{section_methods}
we describe the three alternative methods that we propose to test. In
Sections~\ref{subsection_repetition_test},~\ref{subsection_smearing_test}
and \ref{subsection_blocking_test}
we test these methods. We do so not only in the physically relevant $SU(3)$ gauge
theory but also in the $SU(8)$ gauge theory so as to see how these methods might work in
the theoretically interesting large $N$ limit. We finish with some conclusions.

\section{Gauge fields on a lattice}
\label{section_lattice} 

We work on hypercubic lattices of size $L_s^3L_t$ with lattice spacing $a$ and with
periodic boundary conditions on the fields. Our fields are $SU(N)$ matrices, $U_l$,
assigned to the links $l$ of the lattice. If $l$ is the link emanating from $x$ in
the positive $\mu$ direction then $U_l$ can be written more explicitly as $U_{\mu}(x)$ .
The Euclidean path integral is 
\begin{equation}
Z=\int {\cal{D}}U \exp\{- \beta S[U]\},
\label{eqn_Z}
\end{equation}
where ${\cal{D}}U$ is the Haar measure and we use the standard plaquette action,
\begin{equation}
\beta S = \beta \sum_p \left\{1-\frac{1}{N} {\text{ReTr}} U_p\right\}  
\quad ; \quad \beta=\frac{2N}{g^2}\equiv\frac{2N}{g^2_L(a)}.
\label{eqn_S}
\end{equation}
Here $U_p$ is the ordered product of link matrices around the plaquette $p$.
More explicitly we can write $U_p$ as $U_{\mu\nu}(x)$ where $p$ is the
plaquette emanating positively from $x$ in the $\mu\nu$ plane, i.e.
\begin{equation}
U_{\mu\nu}(x)=U_{\mu}(x)U_{\nu}(x+a\hat{\mu})U^\dagger_{\mu}(x+a\hat{\nu})U^\dagger_{\nu}(x).
\label{eqn_U}
\end{equation}
We write $\beta=2N/g^2$ since in this way we recover the usual continuum action when we
take the continuum limit of the lattice theory. Since the fields are all defined on
the scale $a$, this defines a running coupling on the scale $a$ which we denote $g^2_L(a)$,
with the label $L$ reminding us of the running coupling scheme used. Since the theory is
asymptotically free $g^2_L(a)\to 0$ as $a\to 0$ and hence $\beta\to \infty$ as $a\to 0$,
i.e. we approach the continuum limit of the theory by increasing $\beta$.
We use a standard heatbath plus over-relaxation algorithm to generate the gauge fields
appropriate to this action and partition function. We also make sure that we choose
$L_t$ large enough that we are on the confining side of the $SU(N)$ finite temperature
deconfining transition
\cite{LTW-Tc}
and we choose the lattice coupling so as to be on the weak coupling side of the 
strong-to-weak coupling transition. (For the location of the latter in
$SU(N\geq 5)$ see Table 16 of
\cite{LTW-Tc},
while for $SU(4)$ see Fig.1 of
\cite{BLMT_N}
and for $SU(3)$ and $SU(2)$ see, for example, Fig.4a of
\cite{KIGSMT-SU3-1983}
and Fig.8 of
\cite{KIGSMT-SU2-1983}
respectively.)

\section{Topology on a lattice}
\label{section_topology}

A Euclidean $D=3+1$ $SU(N)$ gauge field possesses a topological charge $Q$ which is integer-valued 
in a space-time volume with periodic boundary conditions. This charge can be expressed
as the integral over Euclidean space-time of a topological charge density, $Q(x)$, where
\begin{equation}
Q(x)= \frac{1}{32\pi^2} \epsilon_{\mu\nu\rho\sigma}
\mathrm{Tr}\{F_{\mu\nu}(x)F_{\rho\sigma}(x)\}.
\label{eqn_Q_cont}
\end{equation}
Since the plaquette matrix $U_{\mu\nu}(x) = 1 + a^2 F_{\mu\nu}(x) + ....$ on sufficiently
smooth fields, one can write a lattice topological charge density $Q_L(x)$ on
such fields as
\begin{equation}
Q_L(x) \equiv {\frac{1}{32\pi^2}} \epsilon_{\mu\nu\rho\sigma}
\mathrm{Tr}\{U_{\mu\nu}(x)U_{\rho\sigma}(x)\}
= a^4 Q(x) +O(a^6).        
\label{eqn_Q_lat}
\end{equation}
However this definition lacks the reflection properties of the continuum operator
in eqn(\ref{eqn_Q_cont}), since all the plaquettes $U_{\mu\nu}(x)$ are defined as forward
going in terms of our coordinate basis. So to impose the reflection positivity properties
that are needed when calculating the topological charge of realistically fluctuating fields,
rather than fields that are `sufficiently smooth', 
we use the version of this operator that is symmetrised with respect to forward
and backward directions
\cite{DiVecchia-FFD}:
\begin{equation}
Q_L(x) \equiv {\frac{1}{16\times 32\pi^2}} \tilde{\epsilon}_{\alpha\beta\gamma\delta}
\mathrm{Tr}\{U_{\alpha\beta}(x)U_{\gamma\delta}(x)\}
= a^4 Q(x) +O(a^6).        
\label{eqn_Q_latb}
\end{equation}
where the subscripts run over $\pm 1$ to $\pm 4$ and $\tilde{\epsilon}_{\alpha\beta\gamma\delta}$
is an extension of $\epsilon_{\mu\nu\rho\sigma}$ such that
$\tilde{\epsilon}_{\alpha\beta\gamma\delta}$=$-\tilde{\epsilon}_{-\alpha\beta\gamma\delta}$ etc.

The fluctuations of $Q_L(x)$ are related to the expectation value of the composite operator
$Q_L^2(x)$ whose operator product expansion contains the unit operator
\cite{DiVecchia-FFD},
so these fluctuations are powerlike in $1/\beta$. On the other hand, the physical value of
$Q_L(x)$ is $O(a^4)$ and
hence exponentially suppressed in $\beta$. Thus as $\beta$ increases the fluctuations
in $Q_L(x)$ and $Q_L=\sum_x Q_L(x)$ diverge compared to the underlying physical
values. In addition the composite operator $Q_L(x)$ also receives a multiplicative
renormalisation $Z(\beta)$
\begin{equation}
  \langle Q_L\rangle_{Q=Q_0} = Z(\beta)Q_0
\label{eqn_ZQ}
\end{equation}
and although $Z(\beta\to\infty) = 1$ it turns out that $Z(\beta) \ll 1$  at currently
accessible values of $\beta$
\cite{Pisa_ZQ}.
So the calculated value of $Q_L$ for a lattice field of charge $Q$ will
be $Q_L=Z(\beta)Q \ll Q$ and will be easily swamped by the fluctuations. 
In practice all this means that one cannot extract the topological charge of a typical lattice
gauge field by directly calculating $Q_L=\sum_x Q_L(x)$ on that gauge field. However one notes that
the fluctuations obscuring the value of $Q$ are essentially ultraviolet, while the physically relevant
topological charge is on physical length scales. Thus if we perform a very limited local smoothening
of the fields to suppress the ultraviolet fluctuations, this should not affect physics
on long distance scales, and the value of $Q_L=\sum_x Q_L(x)$ calculated on these smoothened
fields should provide a reliable estimate of $Q$. Moreover, recalling that the total topological
charge of a continuum gauge field is unchanged under smooth deformations, we can expect
that even under a moderately large amount of continued smoothening the value of  $Q_L$ will not
change, even though $Q_L(x)$ itself does gradually change. One convenient way to smoothen the
gauge fields is to locally minimise the action. Such a `cooling' of the original `hot'
lattice gauge field
\cite{MT-cool}
involves sweeping through the lattice one link at a time, precisely like the Monte Carlo
except that one chooses the new link matrix to be the one that minimises the total
action of the plaquettes containing that link matrix. This is a standard technique
that one can find described in more detail in, for example,
\cite{DSMT-Q}.
An alternative smoothening method with perturbatively proven renormalisation properties
is the `gradient flow'
\cite{Luscher:2010iy,Luscher:2011bx,Luscher:2013vga}
which is computationally more demanding and  which has been shown to be numerically
equivalent to cooling
\cite{Bonati:2014tqa,Alexandrou:2015yba,Alexandrou:2017hqw}.
Computationally even more demanding are methods that calculate the topological charge
by counting the number of exact zero modes in, for example, the overlap Dirac operator 
\cite{Neuberger}.
For pure gauge theories one finds
\cite{CTW-2002}
that the occasional difference between this method and cooling is due to the very small
instantons near the lattice cut-off, whose weight in the lattice path integral
is strongly affected by lattice corrections and which do not contribute to the
physical topological charge in the continuum limit.

In this paper we choose to use the `cooling' method to provide
the values of $Q$ and $Q(x)$ against which to compare the results of
the alternative methods that we analyse. For a detailed discussion of this method
we refer to
\cite{DSMT-Q}
and for recent demonstrations of how it works in practice we refer to
\cite{AAMT-SU3,AAMT-SUN}.
Here we merely remind the reader of some relevant features of the
method when applied to gauge fields with a small lattice spacing.
Firstly, the total topological charge, calculated using
eqn(\ref{eqn_Q_lat}), will quickly converge, after only a few cooling
sweeps, to a value that is almost constant and is close to an integer.
We will typically perform up to 20 cooling sweeps. As one cools the lattice,
topological charges of opposite sign will gradually annihilate, so reducing the
action, and if one carries on cooling eventually the remaining net
topological charges will shrink, taking a form close to that
of semiclassical instantons, and each of these will eventually shrink within
a lattice hypercube, leaving behind a gauge singularity, and changing the value of
$Q_L$. (This final step  depends on the lattice action used, but this is what typically
occurs with our plaquette action.) The value of $Q_L$ for a topological charge after cooling
is not exactly an integer, because of tree-level lattice spacing corrections to the charge
that depend on the size of the charge in lattice units, $\rho/a$. For a heuristic
formula for this see, for example, the Appendix of 
\cite{DSMT-Q}.
However as long as the lattice spacing is reasonably small, then the correction is small
enough for there to be no ambiguity in assigning an integer topological charge, as
shown for example in
\cite{AAMT-SU3,AAMT-SUN}.
\section{Methods to be tested}
\label{section_methods} 

Since the main obstacle to extracting the topological charge (density) of a gauge field
arises from the high frequency fluctuations of the fields, the simplest approach is
to suppress these fluctuations, while ensuring that the low frequency fluctuations
that are of physical interest are not significantly affected. We expect this to be
possible because we are in a renormalisable theory. The methods we will investigate
in this paper are the following. In addition two (or  more) of these methods can be used
simultaneously.

\subsection{repetition}
\label{subsection_repetition} 

Suppose we have a lattice gauge field $\{U_l\}_0$ generated at a value of $\beta$
large enough for there to be a large separation between lattice and physical length scales.
If we make $i_h$ further heat bath sweeps through that field we obtain a field  $\{U_l\}_{i_h}$
which differs from  $\{U_l\}_0$ in its highest frequency modes, but is nearly identical
on physical length scales, as long as the number of sweeps, $i_h$, is small enough.
In practice we will only consider $i_h\leq 3$. Suppose we produce
an ensemble of $n_r$ such fields by repeating this process with different
random number choices. Label these fields as $\{U_l\}_{i_h}^{i=1,\ldots,n_r}$. Each of
these $n_r$ fields is just $i_h\leq 3$ sweeps from the field $\{U_l\}_0$ and so contains
nearly the same physical topological charge distribution, but differs randomly in its
high frequency fluctuations. If we calculate $Q^{i=1,\ldots,n_r}_{L,i_h}(x)$ on each of these fields
and average
\begin{equation}
\overline{Q}_{L,i_h}(x)=\frac{1}{n_r}\sum_{i=1}^{n_r} Q^i_{L,i_h}(x)
\label{eqn_Qrep}
\end{equation}
we should have a measure of the topological charge density where the effect of high frequency
modes is reduced. (We assume we have made $n_r$ large enough that we can drop that label.)
Clearly this reduction will become more effective as $i_h$ increases, but in doing so one
also increases the risk of encroaching upon the physical length scales that we wish to
leave unchanged, and so we keep $i_h$ very small. Note that this method is designed
to reduce the additive fluctuations in $Q_L$ and not the multiplicative renormalisation
factor $Z(\beta)$ which is defined, as in eqn(\ref{eqn_ZQ}), when we take the average
over the whole ensemble of lattice fields of charge $Q$.

A brief comment on what happens as we increase the number of sweeps $i_h$ from a lattice field.
If we make $i_h$ large, but not so large that the total
charge $Q$ has a significant chance of changing, the averaged topological charge density will
eventually become translation invariant. That is to say, it becomes completely smooth. Thus
we can expect that even a small number of sweeps will begin to smoothen the average density
and hence we need to keep $i_h$ small. It is also important that each field of the generated
ensemble continues to have the charge $Q$. For sufficiently small $a(\beta)$ this will be
the case because the number of heatbath sweeps between changes in $Q$ grows very rapidly
either as $a(\beta)$ decreases at fixed $N$ or as $N$  increases at fixed $a(\beta)$.
Estimates of this quantity can be found in Table 42 of
\cite{AAMT-SUN}
As we see, in the case of $SU(3)$ one should not use this method for $\beta < 6.235$.
We shall therefore perform our tests of the method at $\beta = 6.235$ and $\beta = 6.50$
as well as in $SU(8)$ at $\beta = 46.70$ which does not appear in the Table since there
were no observed changes in $Q$ during the sequence of $O(2\times 10^6)$ heat bath sweeps.
The fact that in $SU(3)$ we need to go to quite high values of $\beta$ is not a real
constraint, since a calculation of the topological charge density will only provide
useful information if the space-time resolution is fine enough, i.e. if  $a(\beta)$
is small enough.

\subsection{smearing}
\label{subsection_smearing} 

Another way to reduce the high frequency fluctuations is to average over neighbouring
fields. On the lattice the fields are unitary matrices $U_l$  on the links $l$. We can
rewrite $U_{\mu}(x)\equiv U_l $ if $l$ is the link that joins the sites $x$ and $x+a\hat{\mu}$.
To average over `neighboring fields' in a gauge invariant way must mean averaging
$U_l$ with nearby paths that also pass from $x$ to  $x+a\hat{\mu}$. The simplest
version, which we shall employ, is to average each $U_{\mu}(x)$ with its forward and
backward going `staples', and to project this sum to a nearby unitary
matrix $U^{i_s=1}_\mu(x)$: 
\begin{eqnarray}
  U^{i_s=1}_\mu(x) = {\cal{P}}\{
    U_\mu(x) & + & p_s\sum_{\nu\neq\mu} U_\nu(x)U_\mu(x+a\hat{\nu})U_\nu^\dagger(x+a\hat{\mu}) \nonumber \\
    & + & p_s\sum_{\nu\neq\mu}U^\dagger_\nu(x-a\hat{\nu})U_\mu(x-a\hat{\nu})U_\nu(x+a\hat{\mu}-a\hat{\nu})
    \}.
    \label{eqn_Usmear}
\end{eqnarray}
Here ${\cal{P}}$ denotes the projection of the sum of matrices back to a nearby unitary
matrix, so that  $U^{i_s=1}_\mu(x)$ transforms in the same way as $U_\mu(x)$ under a local
gauge transformation, and $p_s$
is a constant that we typically choose to be $p_s=1.0$ since that appears to work well
in practice. (Although our comparison of different choices does not claim to be definitive.)
Having defined the once-smeared fields  $U^{i_s=1}_\mu(x)$ in eqn(\ref{eqn_Usmear}), we can
iterate the procedure, replacing all the $U_l$ on the right of eqn(\ref{eqn_Usmear}) by
the corresponding  $U^{i_s=1}_l$, to give us the twice-smeared fields  $U^{i_s=2}_\mu(x)$,
and so on for higher smearing levels $i_s$. We will also sometimes label the original
$U_{\mu}(x)$ fields as $U^{i_s=0}_\mu(x)$. All this is of course very
similar to the typical smearing algorithm used in constructing operators except that
the latter will usually only involve averaging over spatial paths. 

This simplest version also bears some resemblance to the cooling algorithm. This
resemblance is closest for the $SU(2)$ gauge theory. The simplest version of cooling
is again an iterative algorithm with two main differences from the above smearing.
The first is that one excludes the $U_\mu(x)$ in eqn(\ref{eqn_Usmear}) or
equivalently that one sets $p_s\to\infty$. The second is that where any of the
link matrices on the right hand side of eqn(\ref{eqn_Usmear}) have already been
smeared, then one employs the new smeared matrix rather than the original unsmeared one,
with the same approach at higher smearing levels, thus potentially approaching a different
fixed point in the iterative process. Also, for $SU(N>2)$ cooling is implemented
within the $SU(2)$ subgroups of $SU(N)$ (following the standard Cabibbo-Marinari algorithm)
while in smearing the sum of `staples' involves the full $SU(N)$ matrices of the
previous smearing iteration. Nonetheless, despite these differences, we should not be very
surprised if the results of a few levels of smearing are not far from the results of a few 
cooling sweeps, although the effects of smearing are intuitively more transparent.

\subsection{blocking}
\label{subsection_blocking} 

Another possibility is to `integrate out' the shortest distance fluctuations by blocking
the lattice. A simple way is to use a factor of 2 blocking. Just like smearing, this process
can be iterated so we label the blocked links by their level of blocking as $U^{i_b}_\mu(x)$
with the convention that no blocking corresponds to $i_b=0$. (Note that this differs from
a common convention that no blocking corresponds to $i_b=1$ when performing spatial `blocking'
of fields for glueball operators.) We define our singly-blocked fields as a product of
the singly smeared fields defined in eqn(\ref{eqn_Usmear}), i.e.
\begin{equation}
  U^{i_b=1}_\mu(x) = U^{i_s=1}_\mu(x)U^{i_s=1}_\mu(x+a\hat{\mu}).
    \label{eqn_Ublock}
\end{equation}
This field exists on a link of length $2a$ on the blocked lattice, extending from $x$ to
$x+2a\hat{\mu}$, and is clearly unitary with the desired gauge transformation properties.
We produce such blocked links for all sites $x$, effectively doing so for all the $2^4$
blocked lattices. We can iterate this blocking as follows. To produce doubly blocked link
fields we first smear the singly blocked fields $U^{i_b=1}_\mu(x)$ using
the analogue of eqn(\ref{eqn_Usmear}) to produce fields we label $U^{i_b=1,i_s=1}_\mu(x)$.
Our doubly blocked field is then constructed via
\begin{equation}
  U^{i_b=2}_\mu(x) = U^{i_b=1,i_s=1}_\mu(x)U^{i_b=1,i_s=1}_\mu(x+2a\hat{\mu}).
    \label{eqn_Ublock2}
\end{equation}
and will connect sites $4a$ apart. Again these are defined at every site $x$, i.e. effectively
on all the $2^8$ doubly blocked lattices. This process can be iterated to blocking level $i_b$
where the fields live on blocked links that join lattice sites $2^{i_b}a$ apart. 

\section{Testing the Methods}
\label{section_testing} 

In this section we will test how well the methods introduced above work in practice.
In the case of smearing and blocking it will be useful to start by determining how these methods
work with a field that contains a single classical instanton field and how this varies with
the instanton size. In Appendix A we summarise how we construct a lattice `instanton'. We shall
also see below what happens when one combines some of these methods. We are interested to see,
 firstly, whether the methods can reliably identify the total topological charge of a lattice
gauge field and, secondly, whether they can contribute to identifying the topological charge
density of that field.

\subsection{Repetition and topology}
\label{subsection_repetition_test} 

The goal of this method is to determine the topological structure of individual
gauge fields. We begin by considering several thermalised $SU(3)$ lattice fields generated
at $\beta=6.235$ on $18^326$ lattices. Our goal is to say something about their
topological charge density. In units of the string tension $\sigma$ the
lattice spacing is $a\surd\sigma \simeq 0.149$
\cite{AAMT-SUN}
and the lattice volume is the typical one used in  
\cite{AAMT-SUN}
for calculating  $Q$ by cooling, and for calculating $a^2\sigma$. The 6 lattice fields 
we analyse here have $Q=-1,0,0,1,2,3$, as determined after 20 cooling sweeps. We display in
Table~\ref{table_QLb6.235_cool} how the value of the lattice topological charge $Q_L$
varies with $n_c$, the number of cooling sweeps. The two fields with $Q=0$
differ in that the starred field appears to possess a very small charge that disappears
between 5 and 10 cooling sweeps. The cooling histories of the other charges do not
suggest the presence of such a `near-dislocation'. We note that small instantons are
already quite infrequent in $SU(3)$ and become very rare in $SU(N)$ as $N$ increases.
(See
\cite{AAMT-SUN}
for a discussion.)
Starting with each of these six lattice fields fields we perform 3 heat bath sweeps
(labelled $itr=1,2,3$ respectively) using the same value of $\beta=6.235$ at which
the starting fields were
generated. We repeat this for $n_{rep}=10000$ repetitions, each time from the same
starting field but using different random numbers in the heat bath. Thus we have
produced three ensembles of 10000 lattice fields that are 1,2 or 3 heat bath sweeps from
each of the six thermalised starting fields. Since the value of $a$ is small in physical
units, e.g. $a(\beta=6.235) \sim 0.075 \mathrm{fm}$ if we import
the $QCD$ value $1/\surd\sigma \sim 0.5 \mathrm{fm}$ to the pure gauge theory, we expect
that these ensembles will differ from their respective starting fields only at scales
small compared to the physical length scales. So measuring the topological charge
density $Q_L(\vec{x})$ on each of the fields of one of these ensembles, and
averaging the result, should produce a distribution $\bar{Q}_L(\vec{x})$
that encodes the physical topological fluctuations of the starting field while
averaging out its obscuring short distance fluctuations.

We start with the average total topological charge $\bar{Q}_L^{itr}$ for each of the 3 ensembles
labelled by $itr=1,2,3$. We display in Table~\ref{table_QLb6.235_rep} the values of these
charges for each of the six starting fields. For comparison we show for each starting field the
value of $Z(\beta=6.235)Q$ with the renormalisation $Z(\beta=6.235)$ taken from 
\cite{AAMT-SUN}.
This is an  appropriate comparison except that the calculation of $Z(\beta)$  effectively involves
an average over an ensemble of starting fields. We see an approximate agreement with the
values of $Z(\beta)Q$ for the $itr=3$ ensemble, and this indicates that after 3 heat bath sweeps
the ultraviolet fluctuations are largely uncorrelated from those of the starting field.
(The only exception is the $Q=0^\star$ ensemble, which reflects the temporary existence
under cooling of a non-zero charge, as we saw in Table~\ref{table_QLb6.235_cool}.)
So we shall confine most of our analysis to the $itr=3$ ensemble. In
Fig.~\ref{fig_Qtr10K_l18b6.235.tex} we show that the average values of $Q_L$ are close
to the  expectation that $\bar{Q}_L|_{Q=Q_0}\simeq Z(\beta=6.235)Q_0$.
(We have excluded the $Q=0^\star$
value which would clearly not fit since the $Q=0$ value one obtains after 20 cooling sweeps
does not reflect the values in the first few cooling sweeps.) This indicates that for
the typical $SU(3)$ lattice fields that do not contain narrow instantons, the value
of $\bar{Q}_L^{itr=3}$ accurately reflects the value of $Q$ one obtains after 10 or 20
cooling sweeps. We also note that given the small errors
on the values of $\bar{Q}_L^{itr=3}$, as shown in Fig.~\ref{fig_Qtr10K_l18b6.235.tex},
the assignement of a value of $Q$ on the basis of the value of $\bar{Q}_L^{itr=3}$ is
unambiguous. It is interesting to see at what point the connection becomes ambiguous
as we reduce the number of repetitions. As an example we show in 
Fig.~\ref{fig_Qtr1K_l18b6.235.tex} the 10 values that we obtain for $\bar{Q}_L^{itr=3}$
with each value coming from an ensemble of 1000 repeated fields. We see that at the
margins we are beginning to have some ambiguity. So 1000 repetitions is probably
the minimum value one might use to determine $Q$ from  $\bar{Q}_L^{itr=3}$.

Of course the above is an expensive way to determine the total charge $Q$. What we are more
interested in is the possibility of saying something about the distribution of topological
charge across the lattice. As a first step we shall consider its profile in time
\begin{equation}
  Q_L(t) = \sum_{\bar{x}} Q_L(\bar{x},t).
    \label{eqn_Qproft}
\end{equation}
We begin by showing in Fig.~\ref{fig_Qprofcoolt_l18b6.235c} the profile obtained
from our $Q=-1$ field after various numbers of cooling sweeps. As expected cooling rapidly
erases the fluctuations in the topological charge density, even if the total charge is
quasi-stable, and it is not clear at which of the earliest cooling steps
one has a reliable representation of the physical
topological charge density of the field. We now show in Fig.~\ref{fig_Qproft_r10Kl18b6.235}
the average profile we obtain from the ensemble of $10^4$ fields obtained by performing
3 heat bath sweeps starting with the given $Q=-1$ field. We see that the statistical errors are 
very small compared to the fluctuations of $\bar{Q}_L(t)$ with $t$. In fact one finds that
using a smaller ensemble of  $\sim 10^3$ is quite adequate from this point of view,
as shown in Fig.~\ref{fig_Qprofrep_l18b6.235b} where, for the sake of variety, the profile is
in $x$ rather than in $t$. In both plots we also display the profile obtained after
$n_c=2$ cooling sweeps. (Note that in each of the plots both
distributions have been renormalised to a common value of the topological charge $Q_L$
so as to account for renormalisation factor $Z(\beta)$.) We see a striking level of agreement 
between the two distributions in each plot. This would no longer be the case if we had used the
$n_c=5$ profile. This provides some evidence that at this $\beta$ the profile after two
cooling sweeps gives us a reasonably reliable picture of the original field's topological
charge distribution. Of course this is only one example, so we provide in  
Fig.~\ref{fig_Q3proft_r10Kl18b6.235} another, this time using the $Q=3$ thermalised field
as our starting field for the repetitions. All this provides some promising evidence
that such a `repetition' procedure can be useful in practice. The comparison with cooling
also provides some evidence that 2 cooling sweeps on the starting field provides a lattice
field whose topological structure reasonably aproximates that of the starting field.

Of course we would like to check how all this carries over to larger volumes, and to smaller
lattice spacings and to larger $N$. So in Fig.~\ref{fig_Q1proft_c2vsr10Kl26b6.235} we plot
$\bar{Q}_L(t)$ as obtained from a $Q=+1$ starting field on a larger $26^326$ lattice
generated at the same coupling $\beta=6.235$. Once again we see that the field after two
cooling sweeps closely tracks the distribution we have obtained. The main difference
with the smaller volume is that the statistical errors, as compared to the value of
$\bar{Q}_L(t)$,  are slightly larger here. For
asymptoticaly large space-time volumes we would expect that summing over the spatial volume
$V_s$ at fixed $t$ would produce short-distance fluctuations $\propto \surd{V_s}$
and that summing the physical topological fluctuations would also produce 
$\bar{Q}_L(t)\propto \surd{V_s}$ so that the error to signal ratio is weakly dependent
on the volume. So clearly our volumes are not yet asymptotic. As for what happens at
smaller lattice spacings, we show in Fig.~\ref{fig_Q1proft_c2vsr10Kl26b6.500}
the profile obtained for a $Q=-1$ starting field that was generated on a  $26^338$ lattice
at $\beta=6.500$, where the lattice spacing is $a\surd\sigma \simeq 0.104$ in physical
units, i.e. about $1/3$ smaller than at $\beta=6.235$. We clearly obtain a statistically
accurate profile and the profile after 2 cooling sweeps of the starting field is quite
similar, although the level of agreement appears to be slightly less than at $\beta=6.235$.
(The profiles after 1 or 3 cooling sweeps differ much more.) To see what happens at larger
$N$ we analyse a $Q=+1$ lattice field in $SU(8)$, on a $16^324$ lattice at $\beta=46.70$,
corresponding to a lattice spacing $a\surd\sigma \simeq 0.166$, i.e. close to the value
at $\beta=6.235$ in $SU(3)$. We show the resulting profile in 
Fig.~\ref{fig_Q1proft_c2vsr10Kl16b46.70}. Again we see a usefully accurate profile
that is quite closely tracked by the profile obtained after 2 cooling sweeps of the starting
field. Thus we conclude that such `repetition' provides a promisingly precise  method for
elucidating the topological charge density of individual $SU(N)$ lattice gauge fields,
and so can provide a useful benchmark against which to test other candidate methods,
perhaps much more economical, such as cooling.

In utilising, as above, repetitions of 3 heat bath sweeps we assume that performing 3 heat
bath sweeps is very unlikely to alter the topological charge. This was motivated by the
data in Table 42 of
\cite{AAMT-SUN}
for the mean number of heat bath sweeps, $\tau_Q$,  needed to produce a change in $Q$.
In the case of our calculations at $\beta=6.235$ one has $\tau_Q\sim 100$ which while
quite large is not so large that it is obvious that there will not be some attempted
tunnellings in an ensemble of $10^4$ repetitions of 3 heat bath sweeps from a given
starting field. To quantify this issue we have calculated the values of $Q_L$ for each
of the $10^4$ lattice fields in the ensemble, after cooling each of the $10^4$ lattice fields
with 2 sweeps, where the cooling locally minimise the action. (We choose 2 cooling sweeps
because our above calculations have indicated that this should provide a reasonably reliable
representation of the underlying physical topological charge distribution.) We do this
for each of our 5 starting fields, and we plot the 5 histograms (each with $10^4$ entries)
in Fig.~\ref{fig_plot_QLhist_c2tr3_l18b6.235}. The peaked and smoothly decreasing
histograms that we see for the ensembles corresponding to the $Q=2$ and $Q=3$ starting
fields are consistent with the fields in the ensemble having the corresponding charges.
On the other hand the histograms of the other 3 ensembles possess small secondary shoulders
that indicate that a fraction of the fields, at the percent level, have changed $Q$ by unity.
While this shows that these calculations possess some undesirable `background' at this
value of $\beta$ in $SU(3)$, it is reassuring that this `background' is already very
small, and so we can be confident that it will be negligible at significantly smaller
values of $a(\beta)$ in $SU(3)$ and, even more so, at higher $N$.

\subsection{Smearing and topology}
\label{subsection_smearing_test} 

We now turn to assessing the usefulness of the iterative smearing technique described
in Section~\ref{subsection_smearing}. The hope here is that a modest number of smearing
steps applied to an $SU(N)$ gauge field will expose its physical topological charge,
and its space-time distribution, by suppressing the unwanted high frequency fluctuations
while simultaneously leaving the longer distance fluctuations largely undisturbed. 

As a preliminary step we consider classical lattice instanton fields, as constructed in
the Appendix, and we see how their topological charge densities are affected by smearing.
Here there are, of course, no ultraviolet fluctuations, unless the instanton size is so small
that it itself is effectively an ultraviolet fluctuation, and so we are testing how long
distance fluctuations respond to smearing using our particular smearing algorithm. We consider
instantons of various sizes $\rho$. We calculate the total lattice charge $Q_L$ at
each smearing iteration, and plot the values in Table~\ref{table_QL_smearI} for several
values of $\rho$. For comparison we also show the values of $Q_L$ obtained after 8
and 20 cooling sweeps on the original instanton fields. We observe that the instanton
fields are little changed under the moderate number of smearing steps displayed,
except for the smallest, near-ultraviolet instantons with $\rho\leq 1.5$.
As for the topological charge density, we deliberately take as an example a quite
small $\rho=2a$ instanton. We compare in Fig.~\ref{fig_plot_Iprofsmear_r2l18}
the profile $Q_L(t)$, as defined in eqn(\ref{eqn_Qproft}), of the instanton field
after 20 smearing steps with that of the original field. We see only a slight change
even after this quite large number of smearing steps. It is interesting to compare
this to the profiles one obtains after 16 and 20 cooling sweeps of the same 
$\rho=2a$ instanton, which we do in Fig.~\ref{fig_plot_Iprofcool_r2l18}. We see
a pronounced shrinkage after 16 cooling sweeps, and the instanton is in the
process of completely disappearing after 20 cooling sweeps. All this provides
evidence that if we perform less than, say, 10 smearing steps at values of $a(\beta)$
where the typical instanton size is substantially larger than $2a$, then the
physical topological structure should be reliably encoded in the smeared fields.
Of course when we smear a typical lattice field  that has fluctuations on all
length scales, we will want to use enough smearing steps that the high frequency
fluctuations are largely erased. Since the suppression of the total topological
charge $Q_L$, as in eqn(\ref{eqn_ZQ}), is largely due to high frequency fluctuations,
we expect the value of $Q_L$ to initially grow with the number of smearings,
and so a plausible criterion for determining a useful number of smearing steps is to
see when $Q_L$ takes a value near its maximum. To establish what this might be,
we show in Table~\ref{table_QL_smearQ} how the average value of $Q_L$ varies
with the number of smearings $i_s$ for an ensemble of fields with $Q=1$
(as determined after 20 cooling sweeps). We consider 3 separate ensembles,
one in $SU(8)$ obtained on $16^324$ lattices at a coupling $\beta=46.70$, and
two in $SU(3)$, with one on $18^326$ lattices at $\beta=6.235$ and another,
at a smaller lattice spacing, on $26^338$ lattices at $\beta=6.50$.
We observe, as and aside, that the $SU(8)$ calculation and the $\beta=6.235$
$SU(3)$ one, which have almost the same lattice spacing in units of the mass gap,
have a very similar reponse to smearing. More importantly, we observe that in all
three cases, the value of $Q_L$ reaches a plateau after about 4 or 5 smearings.
Thus in our examples below we choose to use 5 or 7 smearing steps.

We turn now to the question of how effectively smearing determines the total topological charge
of realistic lattice fields containing fluctuations on all length scales. As a first example
we choose to look at a sequence of $\sim 50000$ $SU(3)$ gauge fields on $36^344$
lattices, generated at $\beta = 6.70$. Here $a\surd\sigma \simeq 0.079$ and the
typical instanton size (after 20 cooling sweeps) is much larger than $2a$.
We calculate the lattice topological charge every 25 Monte Carlo sweeps, so that
we have $\sim 2000$ fields in total. We both iteratively smear these fields and,
separately, cool them. After 20 cooling sweeps the lattice topological charge is
very close to an integer value and we assign the field the corresponding integer
charge $Q$. In Fig.~\ref{fig_QLhistrun_sm8_l36b6.70} we plot the histograms of
the lattice topological charges obtained on fields after 7 steps of smearing,
indicating in each case what is the charge $Q$. We see that the values of $Q_L$
obtained in this way agree very well with the charge $Q$. The very slight ambiguity,
as indicated by the overlapping tails of distributions of different $Q$, is at no
more than the $\sim 0.1\%$ level. Note that this overlap is not necessarily a `systematic'
error of the algorithm: it may well indicate the presence of rare but genuine small
instantons that are erased by 20 cooling sweeps. We repeat the above calculation
at a coarser lattice spacing, $\beta=6.235$, on an $18^326$ lattice that has about the
same volume in physical units since the lattice spacing is roughly twice as large.
The result is plotted in Fig.~\ref{fig_QLhistrun_sm8_l18b6.235}. Once again
the main distribution of $Q_L$ is peaked around the corresponding values of $Q$, but
the secondary tails are now more pronounced, as one would expect, since there should be
more near-ultraviolet instantons at this larger lattice spacing. An additional
question is how smearing works at larger $N$. To investigate this we take some sequences
of $SU(8)$ fields on $16^324$ lattices at $\beta=46.70$. In $SU(8)$ this value of $\beta$
corresponds roughly to $\beta=6.235$ in $SU(3)$ (using the string tension or the mass
gap as the scale for the comparison of the lattice spacings
\cite{AAMT-SUN}).
As in $SU(3)$ we pick fields every 25 Monte Carlo sweeps and we calculate the value of
$Q_L$ after various numbers of smearing steps. We also cool each (unsmeared) field and
assign a value of $Q$ on the basis of the value $Q_L$ takes after 20 cooling sweeps.
We generate in this way  4 ensembles of fields corresponding to the values $Q=-1,0,+1,+2$
respectively.
The resulting plots of the values of $Q_L$ after 7 smearing steps, for these 4 ensembles,
are plotted in Fig.~\ref{fig_QLhistrun_sm8_l16b46.70}. We see that the values of $Q_L$
are strongly peaked near the corresponding values of $Q$, demonstrating that the value
of $Q_L$ after 7 smearing steps provides an alternative and essentially unambiguous
method to cooling for calculating the physical topological charge of the lattice field.
Comparing to the $SU(3)$ histograms in Fig.~\ref{fig_QLhistrun_sm8_l18b6.235} we observe that
in $SU(8)$ we no longer see the small secondary tails that lead to a slight overlap
between the histograms corresponding to neighbouring values of $Q$. This fits in with
the suggestion that those tails in $SU(3)$ are due to very narrow instantons: in $SU(N)$
such small instantons are exponentially suppressed in $N$
\cite{AAMT-SUN}.
It is interesting to see how such histograms of $Q_L$ vary with the number of smearings.
In Fig.~\ref{fig_Q1proft_sm568l16b46.70} we show the histograms of $Q_L$ for 500 $Q=0$
lattice fields generated on $16^326$ lattices at $\beta=46.70$, after 4, 6 and 8
smearing steps. All distributions, which are quite similar, are sufficiently narrow that
there would be no ambiguity in assigning the value $Q=0$ on the basis of any of these
values of $Q_L$.

Having seen that smearing can efficiently expose the total topological charge of a lattice
field, we turn now to testing whether it can also expose something about the underlying
topological charge density. To do so we calculate the profile $Q_L(t)$ defined in 
eqn(\ref{eqn_Qproft}). In Fig.~\ref{fig_Q1proft_sm6vsr10Kl16b46.70} we calculate $Q_L(t)$
for a single $Q=+1$ $SU(8)$ gauge field on an $16^324$ lattice, generated at $\beta=46.70$,
after 5 smearing steps.
We compare this profile to that obtained from an ensemble of 10000 fields each 3 heat bath
sweeps from the given field (with the latter renormalised to a common value of the
total charge $Q_L$). We see that there is a very close correspondence between the
two distributions. It is obviously relevant to ask how well this close correspondence
will survive under a different number of smearing steps. In Fig.~\ref{fig_Q1proft_sm568l16b46.70}
we compare the profiles after 4, 5, and 8 smearing steps and we observe that the
differences are relatively minor. Turning now to $SU(3)$ we plot 
in Fig.~\ref{fig_Q-1proft_sm6vsr10Kl26b6.50} the profile $Q_L(t)$ after 5 smearing steps
of a single $Q=-1$ gauge field on an $26^338$ lattice generated at $\beta=6.50$.
We again compare this to the profile obtained from an ensemble of 10000 fields each
3 heat bath sweeps from the same $Q=-1$ lattice field (and normalised to a common value of $Q_L$).
The densities are again similar. These and similar studies demonstrate that smearing also
provides an efficient method for obtaining useful information about the topological
charge density of lattice gauge fields.

\subsection{Blocking and topology}
\label{subsection_blocking_test} 

Each blocking step increases the lattice spacing by a factor of two, so blocking a classical
instanton field should halve $\rho$ in lattice units. If the blocking algorithm is `sensible'
then one would expect that the value of $Q_L$ one obtains for an instanton that is of size
$\rho$ in the blocked field will not be far from the value of $Q_L$ for an instanton of size
$2\rho$ in the original field. In Table~\ref{table_QL_blockI} we see that this is very much the
case: e.g. a twice blocked $\rho=8$ instanton on a $32^4$ lattice, has an integrated topological
charge that is very close to that of a singly blocked  $\rho=4$ instanton on a $16^4$ lattice, which
in turn is close to that of an unblocked $\rho=2$ instanton on a $8^4$ lattice. This
reassures us that our blocking algorithm is indeed `sensible'. What this also implies is that
blocking a field will produce a narrower instanton whose lattice charge will be smaller,
as we see in Table~\ref{table_QL_blockI}, and so it will be less visible above the statistical
noise, particularly as we would naively expect the fluctuations on the blocked lattice to be 
larger than on the original lattice. This we can hope to compensate for by calculating the
value of $Q_L$ on all the $2^{4i_b}$ different blocked lattices that one can obtain from the
original lattice, and this is what we shall do in our calculations below. That is to say we
calculate blocked links for every site of the original unblocked lattice and use these to
calculate blocked topological charge densities, $Q^{i_b}_L(x)$, at each site of the unblocked
lattice. Of course these have to normalised by the number of blocked lattices,
i.e. by a factor $1/2^{4i_b}$, and this is to be understood in the following calculations.

It is also useful to check that nothing much changes when we block instanton fields that
have been cooled. This we do in Table~\ref{table_QLcool_blockI8} for an instanton of size
$\rho=8a$ on a  $32^4$ lattice. Here the original lattice fields have been cooled by up
to 20 cooling sweeps. We block the lattices after various numbers of cooling sweeps and
calculate the lattice topological charged on those blocked lattices. In
Table~\ref{table_QLcool_blockI8} we list the resulting topological charges.
We observe that while the value of $Q_L$ on the hot $n_c=0$ lattices decreases
with $i_b$, the number of blocking steps, the values on the cooled fields are almost
the same as on the uncooled fields. This is not unexpected: as we see in
Fig.~\ref{fig_Iprofcl_r8l40} a $\rho=8a$ instanton changes very little even after
20 cooling sweeps.

We now test whether blocking a typical Monte Carlo lattice field, with fluctuations on all
length scales, can help us to identify its topological charge. We generate sub-ensembles of fields
of fixed topological charge $Q$ from a long sequence of Monte Carlo generated fields,
where the value of $Q$ is obtained by submitting each field in the sequence to 20 cooling sweeps
and then calculating $Q_L$ on that cooled lattice. This value of $Q_L$ turns out in practice
to be very close to an integer, and we assign to $Q$ this integer value. We block each field up to
three times and calculate  $Q_L$ for the original field and also on all the blocked lattices using
the blocked link matrices. We denote the resulting charge by $Q^{i_b}_L$ where $i_b$ is the
blocking level. In a given subensemble of fields with a given charge $Q$ the values of $Q^{i_b}_L$
will form a distribution with an average value $\langle Q^{i_b}_L \rangle_Q$ and a
half-width spread $\delta Q^{i_b}_L$. If we do no blocking then for typical values of $Q$ one has
$\delta Q^{i_b=0}_L \gg \langle Q^{i_b=0}_L \rangle_Q$ (as is well known and as we shall confirm below)
so for any given lattice field the calculated value of  $Q^{i_b=0}_L$ gives us essentially no
information about the true value of its topological charge. So our question is: does
blocking reduce the fluctuations $\delta Q^{i_b}_L$?

It is clear from the above that this method is most likely to be useful if the lattice spacing
is small enough that the physically relevant topological charges are large enough, in lattice
units, to survive one or two iterations of blocking.
So we will begin in  $SU(3)$ with lattice fields generated at $\beta=6.70$. At this $\beta$
the lattice spacing is $a\surd\sigma = 0.0790(3)$ in units of the string tension
\cite{AAMT-SUN}
and the typical size of a topological charge after 20 cooling sweeps is $\rho\sim 7a-8a$,
as we shall see below.  We produce a well-thermalised sequence
of 110000 such $SU(3)$ gauge fields on a $36^344$ lattice, and perform our
blocking and cooling every twenty-fifth field in the sequence, i.e. on a total of 4400
lattice fields. We block each of these fields up to 3 times. We also cool each of the
unblocked fields with 20 cooling sweeps so as to obtain its integer topological charge $Q$.
We thus have subsets of fields with a given $Q$. For example we have 800 lattice
fields with $Q=+1$ and we calculate the value of the lattice topological charge
$Q^{i_b}_L$ when we block these fields $i_b$ times. We do so up to $i_b=3$ which
corresponds to blocked links that join lattice sites that are $8a$ apart.
We do this for all our values of $Q$ which in practice means from $Q=-3$ to $Q=+3$.
For each value of $Q$ we thus obtain a distribution of values of $Q^{i_b}_L$
and from this we obtain an average value $\langle Q^{i_b}_L \rangle_Q$, with
its statistical error, as well as the half-width of the distribution,
$\delta Q^{i_b}_L$. These values are listed in Table~\ref{table_QLb6.70_blockQ}.
In the $i_b=0$ columns we observe the severe suppression of $\langle Q_L \rangle_Q$
on these fully fluctuating lattice fields, as we expect since
$\langle Q_L \rangle_Q = Z(\beta)Q$ with $Z(\beta=6.70)=0.241(28)$
\cite{AAMT-SUN}.
The fact that the half-width of the fluctuations is so large, $\delta Q^{i_b}_L\simeq 5.3$,
tells us that for a given lattice field we would not be able to discriminate even between
$Q=0$ and, say, $Q=10$: so the measure $Q_L$ is almost useless here. However after one
blocking the value of $\delta Q^{i_b}_L$ decreases by about a factor of 8 and by a further
factor of about 3.5 if one blocks once more. The average value $\langle Q^{i_b}_L \rangle_Q$
changes rather little with increasing from $i_b=0$ to $i_b=2$ but beyond that the blocking
is clearly beginning to erase the topological charge. So, after 2 blockings we see that
the size of the fluctuation is only  $\delta Q^{i_b=2}_L \simeq \langle Q^{i_b=2}_L \rangle_{Q=1}$
so that even  if our ability to discriminate $Q$ from $Q\pm 1$ is poor, it is becoming useful
for $Q$ versus $Q\pm 2$, and rapidly becomes very good beyond that. To illustrate this we
show in Fig.~\ref{fig_Qhistbl3_b6.70b} the distribution of $Q^{ib=2}_L$ for fields with
$Q=\pm 1,\pm 3$. (We exclude intermediate values of $Q$ so as not to clutter up the plot.)

The fact that the fluctuation $\delta Q^{i_b=2}_L$ is largely independent of $Q$ tells us
that the fluctuations in $Z(\beta)$ are relatively unimportant. It is interesting to
see what happens if we calculate the values of $\delta Q^{i_b}_L$ and $\langle Q^{i_b}_L \rangle_{Q}$
when we cool the original lattice fields and block these cooled lattices. We show the results
of this calculation for the subensemble of fields with $Q=1$ in Table~\ref{table_QLcoolb6.70_blockQ1}.
We first note that the values of $\langle Q^{i_b}_L \rangle_{Q=1}$ after $n_c=20$ cooling sweeps
track approximately the values we obtained after 20 cooling sweeps of our $\rho = 8a$ instanton,
as shown in Table~\ref{table_QLcool_blockI8}, providing some confirmation that the typical size
of a physically relevant topological charge at $\beta=6.70$ is indeed not far from $\rho \sim 8a$.
We also note the dramatic decrease with cooling of the fluctuations in $Q_L$ for the unblocked
$i_b=0$ lattices. This is of course well known, and is why cooling is such an efficient method
for calculating the total topological charge, although it does so at the price of erasing most
of the other topological structure of the fields. By contrast we expect the blocking to
maintain the structure on scales of the blocking (or larger) while integrating out the structure
on smaller scales. We note that the fluctuations on the blocked lattices do not decrease rapidly,
if at all, as we increase the number of cooling sweeps on the original lattices. There are a number
of plausible reasons for this, that we will not digress upon here. We merely note that by the
time $n_c=20$ the fluctuations $\delta Q^{i_b}_L$ for $i_b=2$ and $i_b=3$ are so small that
there is no ambiguity in assigning a unique value of $Q$ from the calculated value of $Q^{i_b}_L$.
(Although not shown here, the fluctuations are in fact almost independent of the value of $Q$.)
Indeed this is also largely true even for $n_c=12$. 

Because the blocking increases the scale by a factor of 2 every time it is applied, it will
erase physically interesting fluctuations unless the lattice spacing is sufficiently small.
That is why we applied it above to fields generated at $\beta=6.70$ which is, for example,
the smallest $SU(3)$ lattice spacing used in 
\cite{AAMT-SUN}.
In Table~\ref{table_QLb6.235_blockQ} we show the results of the same type of calculation
as as in Table~\ref{table_QLb6.70_blockQ} except at $\beta=6.235$ which corresponds
to a lattice spacing of $a\surd\sigma = 0.1490(6)$, i.e. almost twice that at $\beta=6.70$.
Comparing Tables~\ref{table_QLb6.235_blockQ} and \ref{table_QLb6.70_blockQ} we see that
the results after a single blocking at $\beta=6.235$ are very much like the results after
two blockings at $\beta=6.70$. Higher blocking levels rapidly become less useful. It
appears from this that the blocking method will cease to be useful if we decrease
$\beta$ significantly below  $\beta=6.235$.

\section{Conclusions}
\label{section_conclusion} 

In this paper we have tested some old suggestions 
\cite{MT-methods}
for alternative methods to calculate the total topological charge, and its distribution in space-time,
of individual lattice gauge fields that are typical of the important fields in the path integral
-- that is to say, ones that possess fluctuations on all length scales. As a benchmark for
comparison in these tests we use 20 sweeps of the efficient and well-known cooling method to obtain
the topological charge of the lattice fields of interest, although it is much less clear how to obtain
a reliable topological charge density using cooling. In practice in our tests we do not examine the full
4 dimensional density but instead calculate profiles in, say, $t$ by summing the charge at each fixed $t$.
Ideally one would like to have an economical method for calculating the profile or density of
a lattice field since comparisons with theoretical expectations may well involve
transforms or correlators of the density, taken over a large sample of lattice fields.

The first method consists of producing an ensemble of gauge fields that are a very few Monte Carlo
sweeps from the given gauge field. One then averages the topological charge and its density over the
fields in this ensemble so as to largely erase the troublesome additive short-distance fluctuations.
We tested this algorithm using 3 Monte Carlo sweeps and an ensemble of up to $10^4$ fields. This is
clearly an expensive method but it has the advantage that there is no conceptual ambiguity: it is clear
that once the lattice spacing is small, 3 Monte Carlo sweeps can have at most a small effect on the
physically relevant long distance fluctuations. We find that for $SU(3)$ such averages produce a
topological charge that reproduces the cooled charge with essentially no ambiguity once the lattice
spacing is small, and with no ambiguity in $SU(8$. In fact we have good evidence that the rare
mismatch in $SU(3)$ is due to the presence of a very small instanton that is erased by 20 cooling
sweeps. More interesting is that one also obtains the topological charge profile with very good
statistical accuracy. Thus it provides a useful benchmark against which to compare other
potential methods for calculating such a profile. In particular if we comparing this result
to the profile that one obtains under cooling, one sees that
typically the profile after 2 cooling sweeps matches quite well in both $SU(3)$ and $SU(8)$,
and so can be used as a much more economical way of calculating the density profile. In other
circumstances a different number of cooling sweeps may be appropriate, but this can be
determined in the same way as we have done here. 

The second method consists of an iterative `smearing' of the fields on the links of the lattice.
This proves to be an economical and reliable method of calculating both the total
charge and its profile. With only around 5 smearing iterations one produces a charge profile
that accurately matches the charge profile one obtains with the first method which uses
an ensemble of $10^4$ fields.
This method has some technical similarities to cooling but is conceptually more transparent.

The third method uses iterative blocking of the fields. This method can also work, but
because it doubles the lattice spacing at each iteration, it can only be useful for
lattice fields that have an extremely small lattice spacing, which makes it much less
useful for current lattice simulations.

All three methods can be altered in various ways and indeed could be used in various combinations.
Also one could use an `improved' lattice topological charge, as for example in
\cite{QL_imp,QL_imp_oz},
which aims to remove the leading $O(a^6)$ correction in the relation $Q_L(x)=a^4Q(x) + O(a^6)$.
We have not attempted to optimise these methods in such ways, but the fact that the simplest
approaches are already quite successful suggests that some further optimisation might
prove worthwhile.

\section*{Acknowledgements}

This work was supported by Oxford Theoretical Physics and All Souls College. The numerical calculations were performed on the Oxford Theoretical Physics cluster.

\appendix

\section{Instanton on a lattice}
\label{appendix_A}

To construct an approximate instanton on a finite periodic lattice we proceed as follows.
We begin by constructing an instanton field in the continuum in an infinite volume but
with fields that vanish at infinity, i.e. $A_\mu(x)\stackrel{x^2\to \infty}{\longrightarrow}0$
where $x^2= x_{\mu}x_{\mu}$.
This provides us with an approximate instanton at the centre of a finite periodic volume when
that volume
is much larger than the size of the instanton. We then discretise the field onto a hypercubic
lattice. We consider an instanton in $SU(2)$ since an instanton in $SU(N)$ can be trivially
obtained by embedding the $SU(2)$ instanton into an $SU(N)$ field.

In an infinite volume a standard gauge potential for an instanton of size $\rho$
centered at $x^2=0$ is
\begin{equation}
  \tilde{A}^I_{\mu}(x) = \frac{x^2}{x^2+\rho^2} g^{-1}(x)\partial_{\mu}g(x)
 \quad ; \quad g(x)=\frac{x_0+ix_j\sigma_j}{(x_{\mu}x_{\mu})^{1/2}}
    \label{eqn_Icont}
\end{equation}
with the action independent of $\rho$, as dictated by the classical scale invariance.
Although this field becomes pure gauge
as $x^2/\rho^2\to\infty$, it remains nontrivial since 
the gauge function $g(x)$ winds once around the $SU(2)$ group as we move across the surface
at infinity. To fit this instanton field into a periodic volume which initially we
take to be arbitrarily large, we perform a gauge transformatiom $g^{-1}(x)$ so that the
transformed gauge potential $A^I_{\mu}(x)$ becomes trivial, $A^I_{\mu}(x)\to 0$,
as $x^2\to\infty$ in all directions.
This gauge transformation is clearly singular as $x^2\to 0$ (although the field strengths
are not) and the winding of the gauge
potential now takes place across the infinitesimal hypersphere enclosing the singularity
at $x^2=0$ rather than the hypersphere at infinity. If we now make the periodic volume
finite but much larger than $\rho$, and place the instanton near the centre of that
volume then there will be a slight mismatch at the boundary
which, if one wishes, one can remove by a suitable transformation of the fields. 
Clearly our instanton is only approximate and in any case imposing a finite volume breaks
the original scale invariance. We can now transfer the field $A^I_{\mu}(x)$ to a lattice field
$U^I_{\mu}(x)$ by choosing, for example,
$U^I_{\mu}(x)={\cal{P}}\left\{\exp\int^{x+a\hat{\mu}}_xA^I_{\mu}(x)dx\right\}$
where the integration path is along the lattice link and ${\cal{P}}$ denotes
path ordering. In practice we divide the link into
several sections, exponentiate the gauge potential at the centre of each section, and then
multiply the resulting matrices (in order) so as to obtain a field on the whole link. For
$SU(2)$ the exponentiation is simple using
$\exp\{i\theta n_k\sigma_k\} = \cos(\theta) + i n_k\sigma_k \sin(\theta)$ where $\vec{n}$ 
is a real unit vector. We place the origin of the instanton at the centre of a hypercube
at the centre of our lattice, since it is convenient for the singularity in the gauge
potential not to be located on a lattice link. The lattice volume is periodic so
there will be a jump in the fields at the boundary, visible in the action and
topological charge densities. This can  be eliminated, if one wishes, by performing
a few cooling sweeps (which can be limited to that part of the lattice that
is far from the instanton core).

\begin{figure}[htb]
\begin	{center}
\leavevmode
\input	{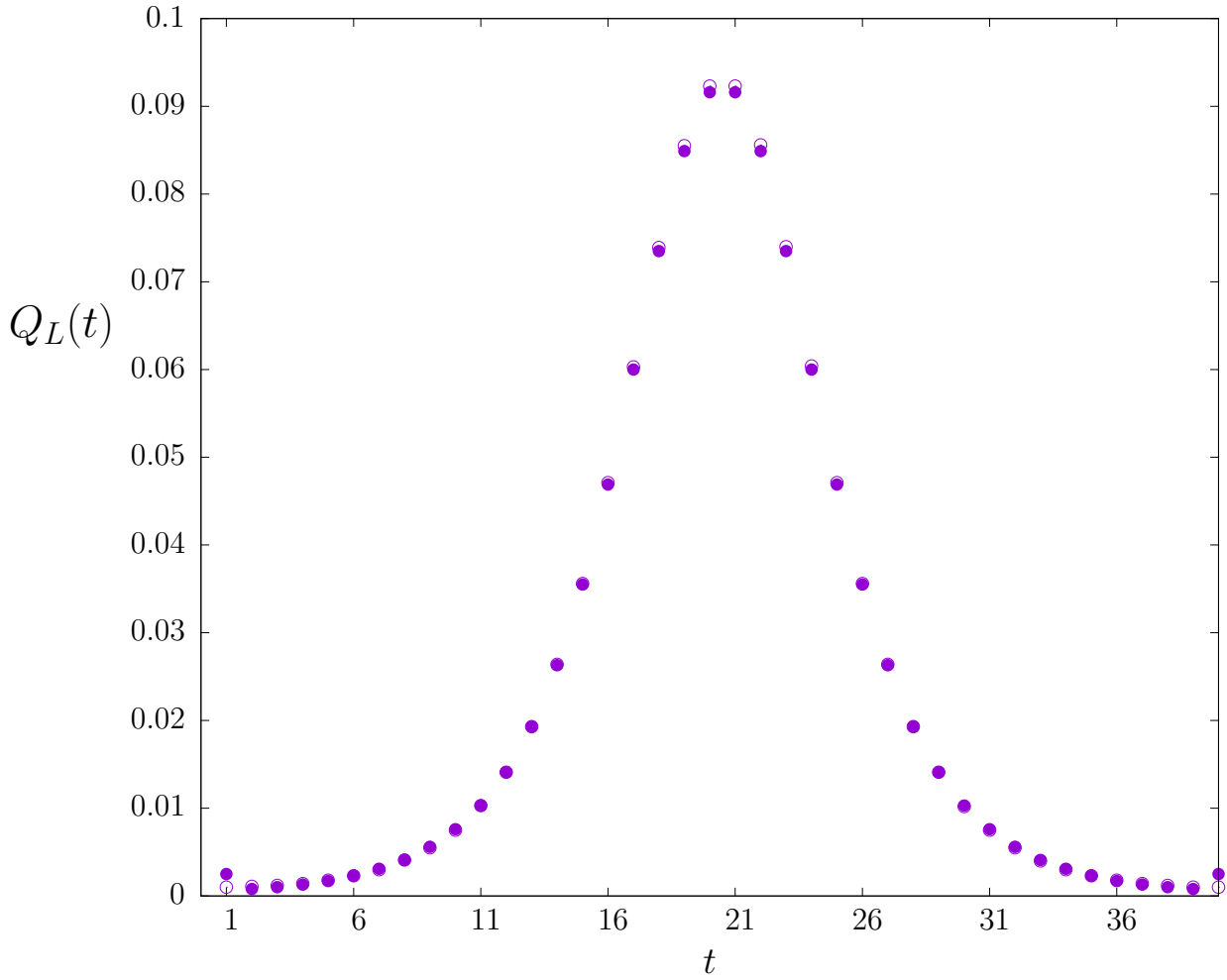}
\end	{center}
\caption{Profile of the integrated lattice topological charge $Q_L(t)$ in each timeslice $t$ of a
  classical instanton with size $\rho=8a$ on a $40^4$ lattice. For the original lattice field,
  $\bullet$, and after 20 cooling sweeps, $\circ$.} 
\label{fig_Iprofcl_r8l40}
\end{figure}

\begin{figure}[htb]
\begin	{center}
\leavevmode
\input	{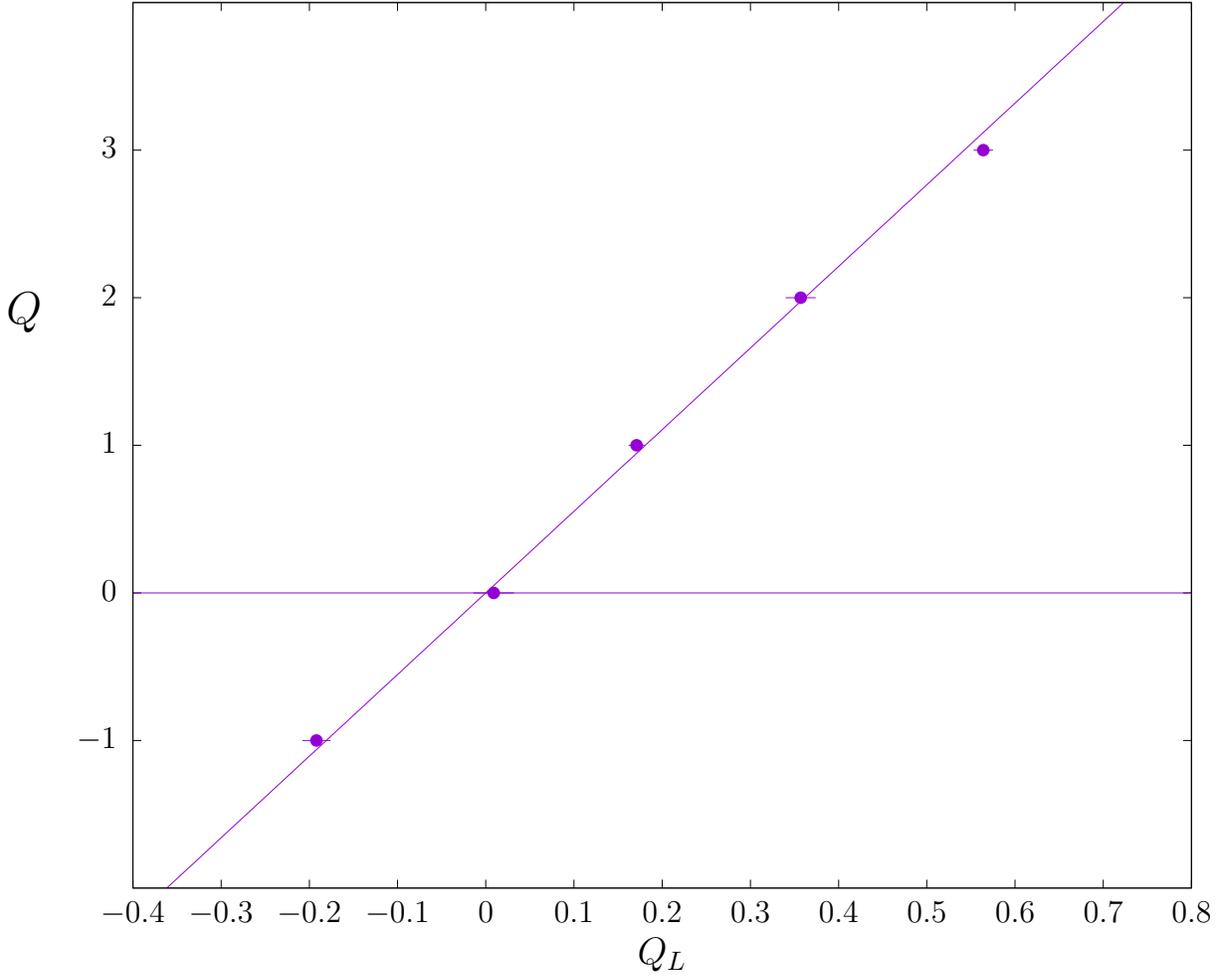}
\end	{center}
\caption{Values of $Q_L$ averaged over 10000 repetitions of 3 sweeps starting from a given $SU(3)$
  field of physical topological charge $Q$, generated at $\beta=6.235$ on a $18^326$ lattice.
  Five separate starting fields with $Q=-1,0,1,2,3$ respectively. Diagonal line is
  $Q_L = Z(\beta)Q$ with $Z(\beta = 6.235) = 0.1808$
  \cite{AAMT-SUN}.}
\label{fig_Qtr10K_l18b6.235.tex}
\end{figure}

\begin{figure}[htb]
\begin	{center}
\leavevmode
\input	{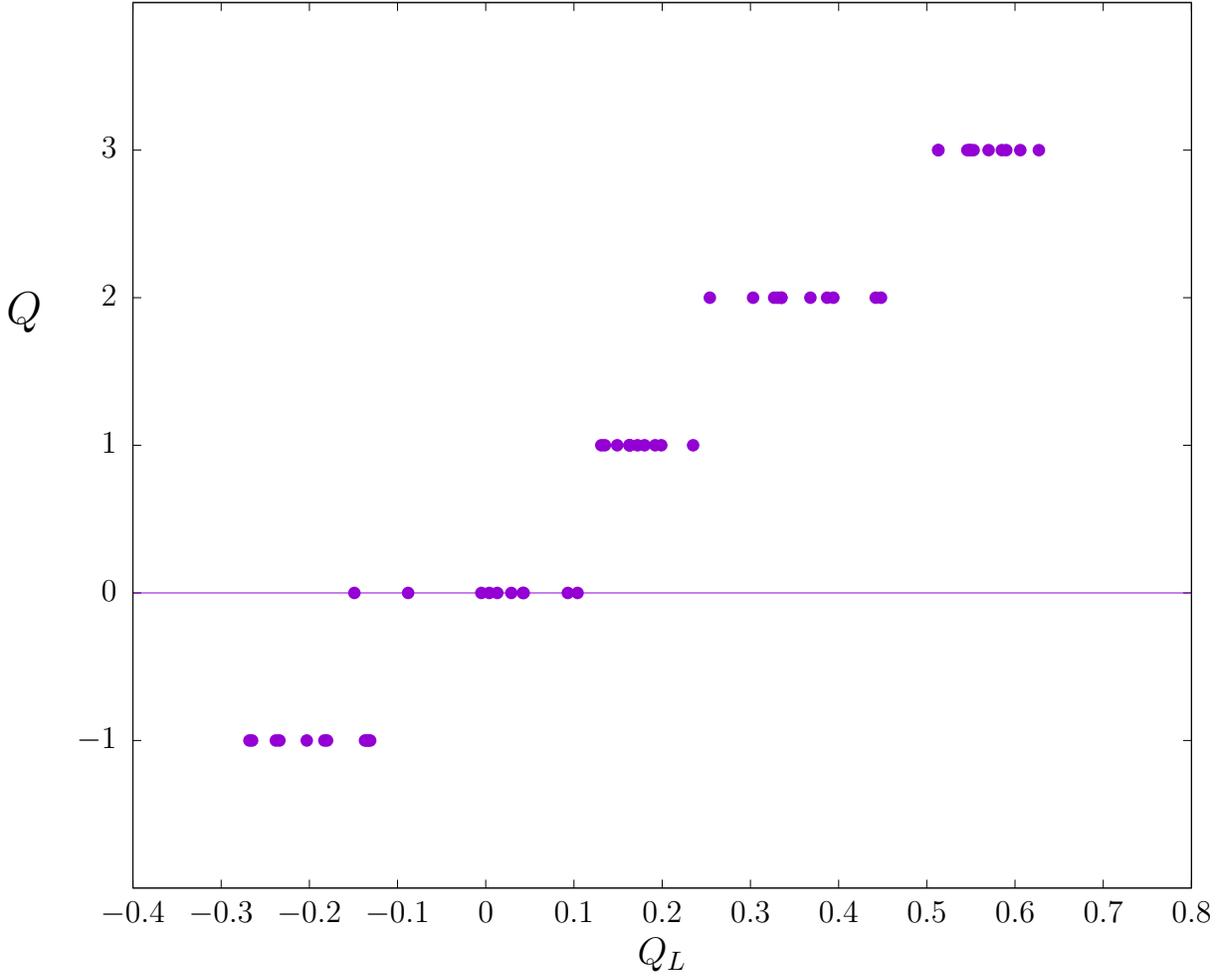}
\end	{center}
\caption{Values of $Q_L$ averaged over 1000 repetitions of 3 sweeps starting from a given $SU(3)$
  field of physical topological charge $Q$, generated at $\beta=6.235$ on a $18^326$ lattice.
  Five separate starting fields with $Q=-1,0,1,2,3$ respectively. With 10 or 11 such averages for
  each starting field.}
\label{fig_Qtr1K_l18b6.235.tex}
\end{figure}

\begin{figure}[htb]
\begin	{center}
\leavevmode
\input	{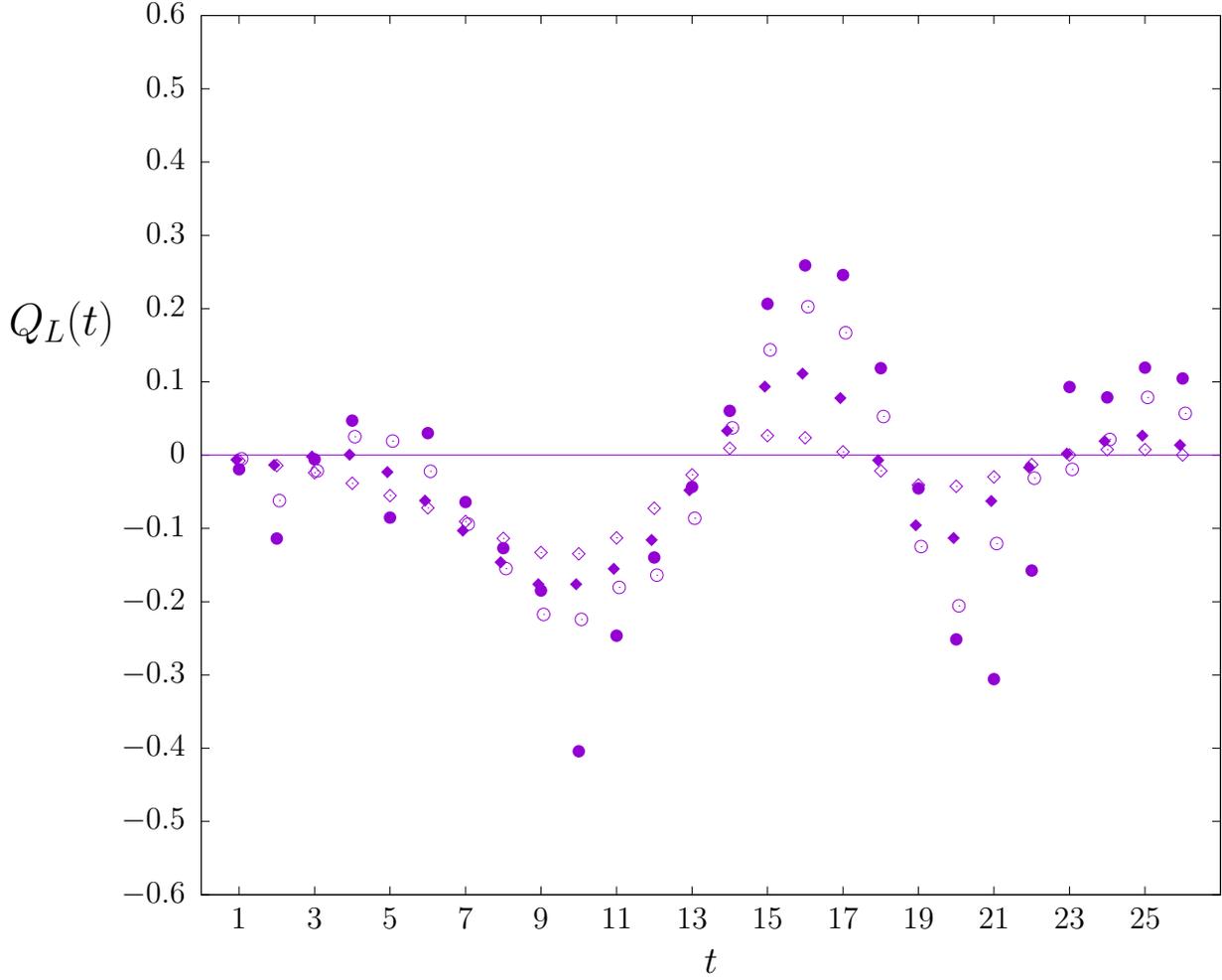}
\end	{center}
\caption{Profiles in $t$ of the lattice topological charge from a $Q=-1$ single $SU(3)$ gauge field on an $18^326$ lattice,
  generated at $\beta=6.235$. Shown are profiles
  after 2 ($\bullet$), after 5 ($\circ$) , after 10 ($\blacklozenge$) and after 20 ($\Diamond$) cooling sweeps.} 
\label{fig_Qprofcoolt_l18b6.235c}
\end{figure}

To see how this construction works we show in Fig.\ref{fig_Iprofcl_r8l40} the topological
charge in each time slice, $Q_L(t)=\sum_{\bar{x}}Q_L(t,\bar{x})$, for an instanton of
size $\rho=8a$ on a $40^4$ lattice centered at $x_{\mu}=20.5$ and constructed
as described above. We also show the profile
one obtains if one performs 20 cooling sweeps starting with this instanton field.
We see that this removes the slight irregularity in the density at $t=1$ and $t=40$
but otherwise leaves the field almost unchanged. In Table~\ref{table_QL_coolI} we
show how the the value of $Q_L$ varies with cooling for instantons of various sizes $\rho$.
We see that under cooling the value of $Q_L$ initially increases slightly (except for the
very smallest values of $\rho$) reflecting the slight imperfection of our instanton
construction. Under further cooling $Q_L$ eventually begins to decrease as the instanton
shrinks, although for the largest values of $\rho$ we see that there is no
visible change in $Q_L$ up to 20 cooling sweeps, which is the number we typically
perform to determine the topological charge of a lattice gauge field. 

\enlargethispage{0.3cm}

If one wants such an instanton in an $SU(N)$ gauge theory one simply takes an
$SU(N)$ field where all link matrices are unit matrices and then replaces the
$2\times 2$ piece at the top left hand corner of the unit matrix by the corresponding
$SU(2)$ instanton matrix. If one wishes one can perform a random gauge transformation
on this $SU(N)$ lattice field. One can also produce multi-instanton fields by
attaching periodic subvolumes to each other containing (anti-)instantons to each other,
in the desired pattern, and performing as needed a few cooling sweeps to smoothen out any slight
discontinuities  at the various boundaries.

\clearpage

%
%

\begin{table}[htb]
\centering
\begin{tabular}{|c|ccccc|} \hline
\multicolumn{6}{|c|}{$Q_L$ of an instanton of size $\rho$ after $n_c$ cools} \\ \hline
$\rho/a$ & $n_c=0$ & $n_c=2$ & $n_c=4$ & $n_c=12$ & $n_c=20$ \\ \hline
8  &  0.979 & 0.987  & 0.987  & 0.987  & 0.987  \\
6  &  0.969 & 0.977  & 0.977  & 0.977  & 0.977  \\
4  &  0.945 & 0.950  & 0.950  & 0.948  & 0.946   \\
3  &  0.911 & 0.912  & 0.911  & 0.906  & 0.901   \\
2  &  0.811 & 0.798  & 0.783  & 0.663  & 0.003   \\ \hline
\end{tabular}
\caption{Integrated topological charge density versus $n_c$ cooling sweeps for our instanton
  field with size $\rho/a=2,3,4,6,8$ on $12^4,16^4,18^4,24^4,32^4$ lattices respectively.}
\label{table_QL_coolI}
\end{table}

%
%

\begin{table}[htb]
\centering
\begin{tabular}{|c|ccccccc|} \hline 
 \multicolumn{8}{|c|}{$Q_L$ cooling histories} \\ \hline
 $Q$   & $n_c=0$ & $n_c=1$ & $n_c=2$ & $n_c=3$ & $n_c=5$ & $n_c=10$ & $n_c=20$ \\ \hline
$-1$   & -1.90 & -0.59 & -0.83 & -0.90 & -0.93 & -0.94 & -0.96  \\
 $0$   & -0.96 &  0.021 & 0.049 & 0.047 & 0.016 & 0.002 &  0.001 \\
 $0^\star$ & 2.89  & 0.87 & 0.53 & 0.54 & 0.39 & 0.072 & 0.065  \\
$+1$   & 1.91  & 0.40 & 0.67 & 0.80 & 0.83 & 0.83 & 0.91  \\
$+2$   & 0.14  & 1.34 & 1.63 & 1.75 & 1.85 & 1.91 & 1.93  \\
$+3$   & 1.22  & 1.93 & 2.45 & 2.60 & 2.74 & 2.82 & 2.86  \\ \hline
\end{tabular}
\caption{Lattice topological charge $Q_L$ after $n_c$ cooling sweeps for each of
  six thermalised $SU(3)$ fields on an $18^326$ lattice, generated at $\beta=6.235$.
  $Q$ is the integer charge inferred from the value of $Q_L$ after 20 cooling sweeps.}
  \label{table_QLb6.235_cool}
\end{table}

\begin{table}[htb]
\centering
\begin{tabular}{|c|ccc|c|} \hline 
 $Q$   & $\bar{Q}_L(itr=1)$ & $\bar{Q}_L(itr=2)$ & $\bar{Q}_L(itr=3)$ & $Z(\beta=6.235)Q$ \\ \hline
$-1$   & -0.165(16)  & -0.170(17) & -0.192(17)  & -0.181  \\
 $0$   & -0.033(25)  &  0.002(22) &  0.009(23)  &  0.0 \\
 $0^\star$ & 0.393(9) & 0.132(14)  & 0.115(20)  &  0.0 \\
$+1$   &  0.165(16)  & 0.135(15)  &  0.171(17)  &  0.181 \\
$+2$   &  0.492(15)  & 0.415(16)  &  0.357(15)  &  0.362 \\
$+3$   &  0.502(18)  & 0.566(16)  &  0.564(16)  &  0.543 \\ \hline
\end{tabular}
\caption{Average lattice topological charge from 10000 fields that
  are {\it itr} sweeps from the given single thermalised field with charge $Q$. On
  an $18^326$ lattice at $\beta=6.235$. For comparison
  we show the value of $Z(\beta=6.235)Q$.}
  \label{table_QLb6.235_rep}
\end{table}

%
%

\begin{table}[htb]
\centering
\begin{tabular}{|c|ccccccc|} \hline
\multicolumn{8}{|c|}{$Q_L(\rho)$ for smeared instanton fields} \\ \hline
$i_s$ & $\rho=6.0a$  & $\rho=4.0a$ & $\rho=3.0a$ & $\rho=2.5a$  & $\rho=2.0a$  & $\rho=1.5a$  & $\rho=1.0a$ \\ \hline
0  &  0.977  & 0.951  & 0.912  & 0.876 & 0.812 & 0.684 & 0.401  \\
1  &  0.978  & 0.951  & 0.913  & 0.876 & 0.809 & 0.662 & 0.284  \\ 
2  &  0.978  & 0.951  & 0.913  & 0.875 & 0.805 & 0.638 & 0.168  \\ 
4  &  0.978  & 0.950  & 0.912  & 0.873 & 0.799 & 0.577 & 0.026  \\ 
6  &  0.978  & 0.950  & 0.911  & 0.872 & 0.793 & 0.486 & 0.003  \\ 
8  &  0.978  & 9,950  & 0.911  & 0.871 & 0.786 & 0.347 & 0.001  \\ 
16 &  0.978  & 0.950  & 0.910  & 0.867 & 0.758 & 0.003 & 0.000  \\   \hline
$n_c=8$  & 0.978  & 0.950 & 0.909  & 0.865 & 0.749 & 0.002 & 0.001 \\
$n_c=20$ & 0.978  & 0.949 & 0.904  & 0.849 & 0.008 & 0.000 & 0.000 \\    \hline
\end{tabular}
\caption{The total lattice topological charge calculated for instantons of size
  $\rho$  where the fields have then been smeared $i_s$ times. Lattice sizes are
  $18^4$ and $32^4$ for $\rho\leq 3.0a$ and  $\rho\geq 4.0a$ respectively. Also
  shown are the lattice topological charges after 8 and 20 cooling sweeps on
  the unsmeared gauge fields.}
\label{table_QL_smearI}
\end{table}

\begin{table}[htb]
\centering
\begin{tabular}{|c|ccccccc|c|} \hline
\multicolumn{9}{|c|}{$\langle Q_L \rangle_{Q=1}$ vs smearing} \\ \hline
          & $i_s=0$     & $i_s=2$   & $i_s=4$    & $i_s=5$  & $i_s=6$  &  $i_s=7$  & $i_s=9$   & $n_c=20$  \\ \hline
 $SU(8)$    &  0.164(47)  &  0.673(8)  & 0.802(3)  & 0.824(2) & 0.835(2) & 0.839(2)  & 0.824(2)  & 0.9638(4)   \\   \hline 
 $SU(3)_a$  &  0.129(57)  &  0.663(17) & 0.795(8)  & 0.819(8) & 0.834(8) & 0.839(7)  & 0.829(7)  & 0.9537(32)   \\   \hline 
 $SU(3)_b$  &  0.200(73)  &  0.736(8)  & 0.837(3)  & 0.856(2) & 0.865(2) & 0.868(2)  & 0.853(2)  & 0.9722(10)  \\   \hline
\end{tabular}
\caption{The average lattice topological charge versus number of smearing iterations $i_s$
  for fields with $Q=1$ in $SU(8)$ on $16^324$ lattices at $\beta=46.70$ and 
  in $SU(3)$ on $18^326$ lattices at $\beta=6.235$ ($a$), and $26^338$ lattices at $\beta=6.50$ ($b$).
   Also shown is the charge  after $n_c=20$ cooling sweeps of the unsmeared fields.}
\label{table_QL_smearQ}
\end{table}

%
%

\begin{table}[htb]
\centering
\begin{tabular}{|c|cccc|} \hline
\multicolumn{5}{|c|}{$Q_L(\rho)$ for blocked instanton fields} \\ \hline
$i_b$ & $\rho=8a$,$L=32$ & $\rho=4a$,$L=16$ & $\rho=2a$,$L=8$ &  $\rho=1a$,$L=4$  \\ \hline
0  &  0.979  &  0.941  &  0.800  & 0.381   \\  
1  &  0.944  &  0.804  &  0.411  & 0.007   \\  
2  &  0.804  &  0.411  &  0.010  & 0.000   \\  
3  &  0.411  &  0.010  &  0.000  & 0.000   \\    \hline
\end{tabular}
\caption{The total lattice topological charge calculated for instantons of size
  $\rho$ on $L^4$ lattices where the fields have been blocked $i_b$ times.}
\label{table_QL_blockI}
\end{table}

\begin{table}[htb]
\centering
\begin{tabular}{|c|ccccc|} \hline
\multicolumn{6}{|c|}{$Q^{i_b}_L$ of a $\rho=8a$ instanton versus $n_c$ cools} \\ \hline
$i_b$ & $n_c=0$ & $n_c=2$ & $n_c=4$ & $n_c=12$ & $n_c=20$ \\ \hline
0  &  0.979  &  0.987  & 0.987  & 0.987  & 0.987  \\
1  &  0.944  &  0.948  & 0.949  & 0.949  & 0.949 \\
2  &  0.804  &  0.805  & 0.807  & 0.807  & 0.806 \\
3  &  0.411  &  0.410  & 0.408  & 0.405  & 0.401 \\  \hline
\end{tabular}
\caption{An instanton of size $\rho=8a$ on a $32^4$ lattice. The field is blocked
  $i_b$ times after first $n_c$ cooling sweeps. Listed is the total lattice topological
 charge $Q^{i_b}_L$.}
\label{table_QLcool_blockI8}
\end{table}

\begin{table}[htb]
\centering
\begin{tabular}{|c|cc|cc|cc|cc|} \hline 
\multicolumn{9}{|c|}{$SU(3) \, , \, 36^344 \, , \, \beta=6.70 \,$ : no cooling} \\ \hline
       & \multicolumn{2}{|c|}{$i_b=0$} & \multicolumn{2}{|c|}{$i_b=1$} & \multicolumn{2}{|c|}{$i_b=2$} & \multicolumn{2}{|c|}{$i_b=3$}  \\ \hline
 $Q$   & $\langle Q^{i_b}_L\rangle_Q$ & $\delta{Q^{i_b}_L}$ & $\langle Q^{i_b}_L\rangle_Q$ & $\delta{Q^{i_b}_L}$ & $\langle Q^{i_b}_L\rangle_Q$ & $\delta{Q^{i_b}_L}$ & $\langle Q^{i_b}_L\rangle_Q$ & $\delta{Q^{i_b}_L}$ \\ \hline
 $0$   & -0.12(14)  & 5.37  & -0.037(23)  & 0.751 & -0.007(8)  & 0.207 & -0.010(5)  & 0.084  \\
$+1$   &  0.41(16)  & 5.38  &  0.272(22)  & 0.699 &  0.199(11) & 0.203 &  0.087(7)  & 0.083 \\
$-1$   & -0.36(24)  & 5.22  & -0.292(29)  & 0.721 & -0.191(7)  & 0.207 & -0.088(5)  & 0.082  \\
$+2$   &  0.22(30)  & 5.35  &  0.531(30)  & 0.704 &  0.405(11) & 0.216 &  0.167(9)  & 0.087 \\
$-2$   & -0.55(18)  & 5.40  & -0.519(26)  & 0.723 & -0.408(10) & 0.216 & -0.173(7)  & 0.088  \\
$+3$   &  0.67(39)  & 5.34  &  0.819(54)  & 0.749 &  0.625(18) & 0.202 &  0.274(11) & 0.093 \\
$-3$   & -0.79(32)  & 5.25  & -0.907(34)  & 0.702 & -0.618(11) & 0.196 & -0.259(9)  & 0.084 \\ \hline
\end{tabular}
\caption{Values of lattice topological charge $Q^{i_b}_L$ on fields blocked $i_b$ times
  from an ensemble of $SU(3)$ fields on $36^344$ lattices generated at $\beta=6.70$, versus
  a charge $Q$ obtained on unblocked fields after 20 cooling sweeps. Also the
  size $\delta{Q^{i_b}_L}$ of the fluctuations around the average value $Q^{i_b}_L$ in
  each case.} 
\label{table_QLb6.70_blockQ}
\end{table}

\begin{table}[htb]
\centering
\begin{tabular}{|c|cc|cc|cc|cc|} \hline 
\multicolumn{9}{|c|}{$SU(3) \, , \, 36^344 \, , \, \beta=6.70 \,$ : cooled $Q=1$ fields} \\ \hline
       & \multicolumn{2}{|c|}{$i_b=0$} & \multicolumn{2}{|c|}{$i_b=1$} & \multicolumn{2}{|c|}{$i_b=2$} & \multicolumn{2}{|c|}{$i_b=3$}  \\ \hline
 $n_c$   & $\langle Q^{i_b}_L\rangle_{Q=1}$ & $\delta{Q^{i_b}_L}$ & $\langle Q^{i_b}_L\rangle_{Q=1}$ & $\delta{Q^{i_b}_L}$ & $\langle Q^{i_b}_L\rangle_{Q=1}$ & $\delta{Q^{i_b}_L}$ & $\langle Q^{i_b}_L\rangle_{Q=1}$ & $\delta{Q^{i_b}_L}$ \\ \hline
 0  &  0.41(16)  & 5.38  &  0.272(22) & 0.70  & 0.199(11) & 0.20 & 0.087(7)  & 0.08 \\
 1  &  0.821(21) & 0.74  &  0.571(16) & 0.46  & 0.332(13) & 0.24 & 0.126(9)  & 0.11 \\
 2  &  0.840(7)  & 0.21  &  0.681(12) & 0.29  & 0.396(13) & 0.22 & 0.144(10) & 0.12 \\
 3  &  0.899(4)  & 0.10  &  0.740(9)  & 0.20  & 0.444(13) & 0.21 & 0.159(11) & 0.13 \\
 8  &  0.959(2)  & 0.022 &  0.855(5)  & 0.074 & 0.585(11) & 0.16 & 0.216(15) & 0.15 \\
20  &  0.978(1)  & 0.013 &  0.917(4)  & 0.045 & 0.717(12) & 0.12 & 0.309(20) & 0.17 \\ \hline
\end{tabular}
\caption{Values of lattice topological charge $Q^{i_b}_L$ obtained on 
  $SU(3)$ lattice fields with topological charge $Q=+1$ generated at $\beta=6.70$,
  then followed by $n_c$ cooling sweeps, and then blocked $i_b$ times. Also the
  size of the fluctuations, $\delta{Q^{i_b}_L}$, around the average value of $Q^{i_b}_L$.}
\label{table_QLcoolb6.70_blockQ1}
\end{table}

\begin{table}[htb]
\centering
\begin{tabular}{|c|cc|cc|cc|cc|} \hline 
\multicolumn{9}{|c|}{$SU(3) \, , \, 18^326 \, , \,  \beta=6.235 \,$ : no cooling} \\ \hline
       & \multicolumn{2}{|c|}{$i_b=0$} & \multicolumn{2}{|c|}{$i_b=1$} & \multicolumn{2}{|c|}{$i_b=2$} & \multicolumn{2}{|c|}{$i_b=3$}  \\ \hline
 $Q$   & $\langle Q^{i_b}_L\rangle_Q$ & $\delta{Q^{i_b}_L}$ & $\langle Q^{i_b}_L\rangle_Q$ & $\delta{Q^{i_b}_L}$ & $\langle Q^{i_b}_L\rangle_Q$ & $\delta{Q^{i_b}_L}$ & $\langle Q^{i_b}_L\rangle_Q$ & $\delta{Q^{i_b}_L}$ \\ \hline
 $0$   & -0.15(7)  & 1.75  & -0.023(10) & 0.25  & -0.006(4)  & 0.083  & -0.001(1) & 0.018  \\
$+1$   &  0.30(7)  & 1.71  &  0.177(11) & 0.25  &  0.084(4)  & 0.082  &  0.009(1) & 0.017  \\
$-1$   & -0.22(6)  & 1.66  & -0.196(9)  & 0.26  & -0.092(4)  & 0.081  & -0.-11(1) & 0.019  \\
$+2$   &  0.38(13) & 1.63  &  0.376(20) & 0.22  &  0.170(8)  & 0.080  &  0.017(2) & 0.019  \\
$-2$   & -0.40(8)  & 1.53  & -0.355(13) & 0.25  & -0.164(5)  & 0.091  & -0.017(1) & 0.020  \\
$+3$   &  0.23(22) & 1.70  &  0.548(37) & 0.23  &  0.242(14) & 0.087  &  0.021(4) & 0.018  \\
$-3$   & -0.61(13) & 1.57  & -0.538(25) & 0.25  & -0.231(8)  & 0.069  & -0.023(2) & 0.077  \\ \hline
\end{tabular}
\caption{Values of lattice topological charge $Q^{i_b}_L$ on fields blocked $i_b$ times
  from an ensemble of $SU(3)$ fields on $18^326$ lattices generated
  at $\beta=6.235$, versus
  a charge $Q$ obtained on unblocked fields after 20 cooling sweeps. Also the
  size $\delta{Q^{i_b}_L}$ of the fluctuations around the average value $Q^{i_b}_L$ in
  each case.} 
\label{table_QLb6.235_blockQ}
\end{table}

\clearpage

\begin{figure}[htb]
\begin	{center}
\leavevmode
\input	{plot_Qproft_r10Kl18b6.235.tex}
\end	{center}
\caption{Profiles in $t$ of the lattice topological charge density from a single $Q=-1$
  $SU(3)$ gauge field on an $18^326$ lattice, generated at $\beta=6.235$. Shown is the  profile 
  obtained from an ensemble of 10000 fields each 3 heat bath sweeps ($\bullet$) from this
  given field, and the profile after 2 cooling sweeps ($\circ$) of
  this given field, normalised to a common value of $Q_L$.} 
\label{fig_Qproft_r10Kl18b6.235}
\end{figure}

\begin{figure}[htb]
\begin	{center}
\leavevmode
\input	{plot_Qprofrep_l18b6.235b.tex}
\end	{center}
\caption{Profiles in $x$ of the topological charge from a single $Q=-1$ $SU(3)$ gauge field on
  an $18^326$ lattice, generated at $\beta=6.235$. Shown is the  profile 
  obtained from an ensemble of 1000 fields each 3 heat bath sweeps from this given field ($\bullet$),
  and the profile after 2 cooling sweeps ($\circ$) of this given field. Normalised to a common value of $Q_L$.} 
\label{fig_Qprofrep_l18b6.235b}
\end{figure}

\begin{figure}[htb]
\begin	{center}
\leavevmode
\input	{plot_Q3proft_r10Kl18b6.235.tex}
\end	{center}
\caption{Profiles in $t$ of the lattice topological charge density from a single $Q=+3$
  $SU(3)$ gauge field on an $18^326$ lattice, generated at $\beta=6.235$. Shown is the  profile 
  obtained from an ensemble of 10000 fields each 3 heat bath sweeps ($\bullet$) from this
  given field, and the profile after 2 cooling sweeps ($\circ$) of
  this given field, normalised to a common value of $Q_L$.} 
\label{fig_Q3proft_r10Kl18b6.235}
\end{figure}



\begin{figure}[htb]
\begin	{center}
\leavevmode
\input	{plot_Q1proft_c2vsr10Kl26b6.235.tex}
\end	{center}
\caption{Profiles in $t$ of the lattice topological charge density from a single $Q=+1$
  $SU(3)$ gauge field on an $26^326$ lattice, generated at $\beta=6.235$. Shown is the  profile 
  obtained from an ensemble of 10000 fields each 3 heat bath sweeps ($\bullet$) from this
  given field, and the profile after 2 cooling sweeps ($\circ$) of
  this given field, normalised to a common value of $Q_L$.} 
\label{fig_Q1proft_c2vsr10Kl26b6.235}
\end{figure}

\begin{figure}[htb]
\begin	{center}
\leavevmode
\input	{plot_Q-1proft_c2vsr10Kl26b6.500.tex}
\end	{center}
\caption{Profiles in $t$ of the lattice topological charge density from a single $Q=-1$
  $SU(3)$ gauge field on an $26^338$ lattice, generated at $\beta=6.500$. Shown is the  profile 
  obtained from an ensemble of 10000 fields each 3 heat bath sweeps ($\bullet$) from this
  given field, and the profile after 2 cooling sweeps ($\circ$) of
  this given field, normalised to a common value of $Q_L$.} 
\label{fig_Q1proft_c2vsr10Kl26b6.500}
\end{figure}

\begin{figure}[htb]
\begin	{center}
\leavevmode
\input	{plot_Q1proft_c2vsr10Kl16b46.70.tex}
\end	{center}
\caption{Profiles in $t$ of the lattice topological charge density from a single $Q=+1$
  $SU(8)$ gauge field on an $16^324$ lattice, generated at $\beta=46.70$. Shown is the  profile 
  obtained from an ensemble of 10000 fields each 3 heat bath sweeps ($\bullet$) from this
  given field, and the profile after 2 cooling sweeps ($\circ$) of
  this given field, normalised to a common value of $Q_L$.} 
\label{fig_Q1proft_c2vsr10Kl16b46.70}
\end{figure}

\begin{figure}[htb]
\begin	{center}
\leavevmode
\input	{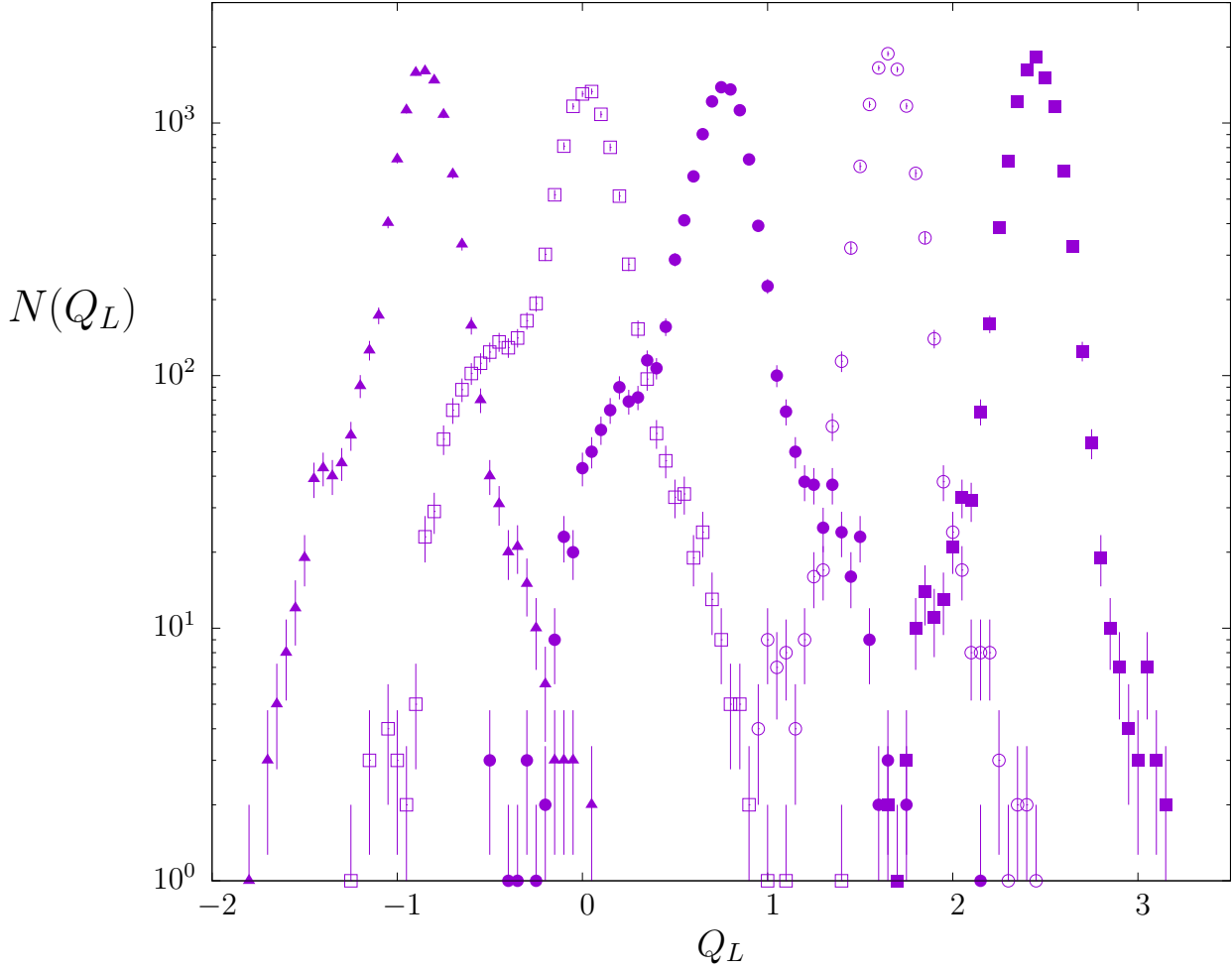}
\end	{center}
\caption{Histogram of lattice topological charges obtained for each lattice field of an
  ensemble of $10^4$ $SU(3)$ fields that are 3 heatbath sweeps plus 2 cooling sweeps from
  a starting field, on a $18^326$ lattice at $\beta=6.235$. Five ensembles are thus analysed,
  from starting fields with $Q=-1$ ($\blacktriangle$), $Q=0$ ($\square$), $Q=+1$ ($\bullet$),
  $Q=2$ ($\circ$) and $Q=3$ ($\blacksquare$).} 
\label{fig_plot_QLhist_c2tr3_l18b6.235}
\end{figure}


\begin{figure}[htb]
\begin	{center}
\leavevmode
\input	{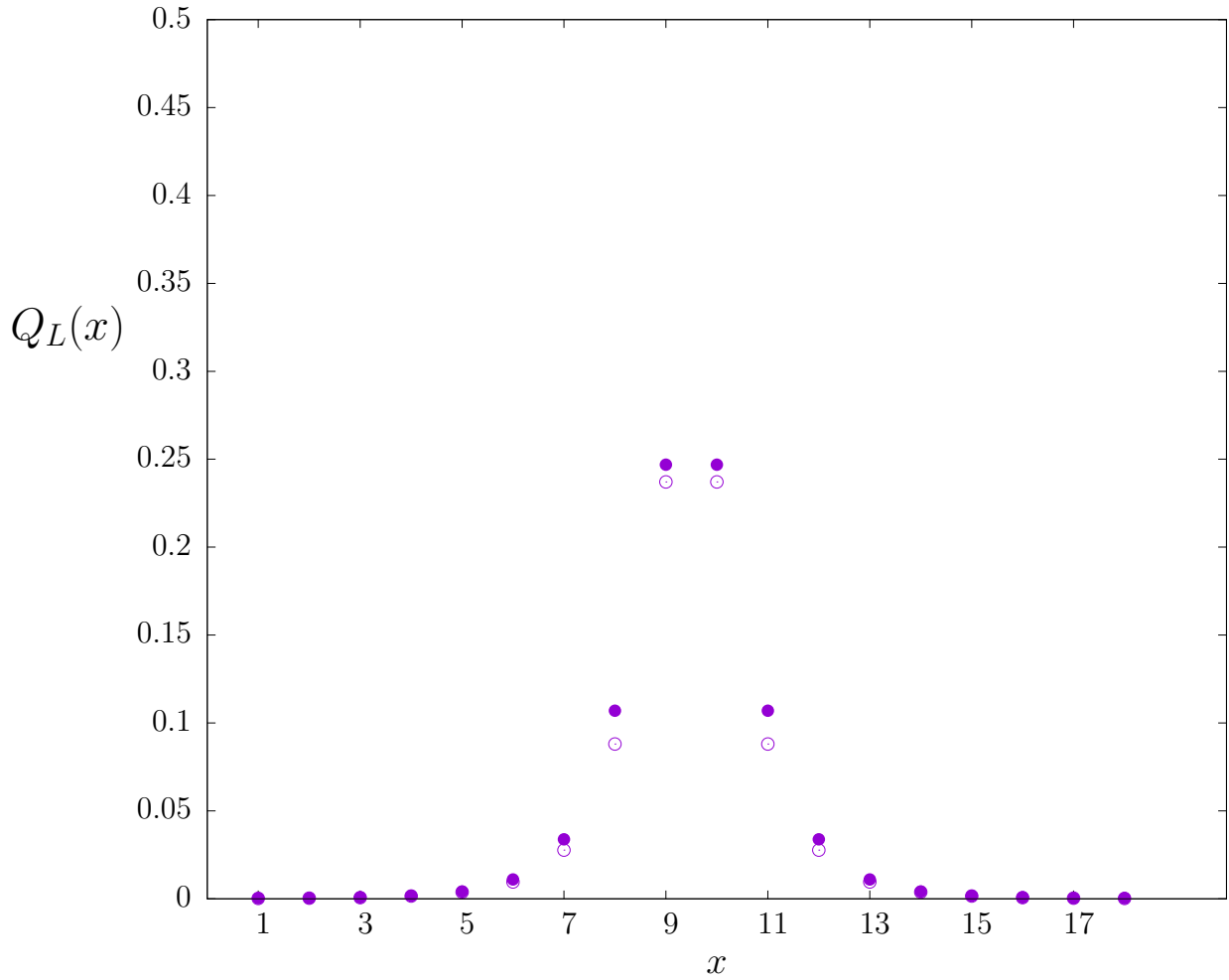}
\end	{center}
\caption{Smearing a classical instanton with $\rho=2a$ on an $18^4$ lattice.
  Profile in $x$ of the topological charge before smearing ($\bullet$) and after 20 smearing iterations ($\circ$).}
\label{fig_plot_Iprofsmear_r2l18}
\end{figure}

\begin{figure}[htb]
\begin	{center}
\leavevmode
\input	{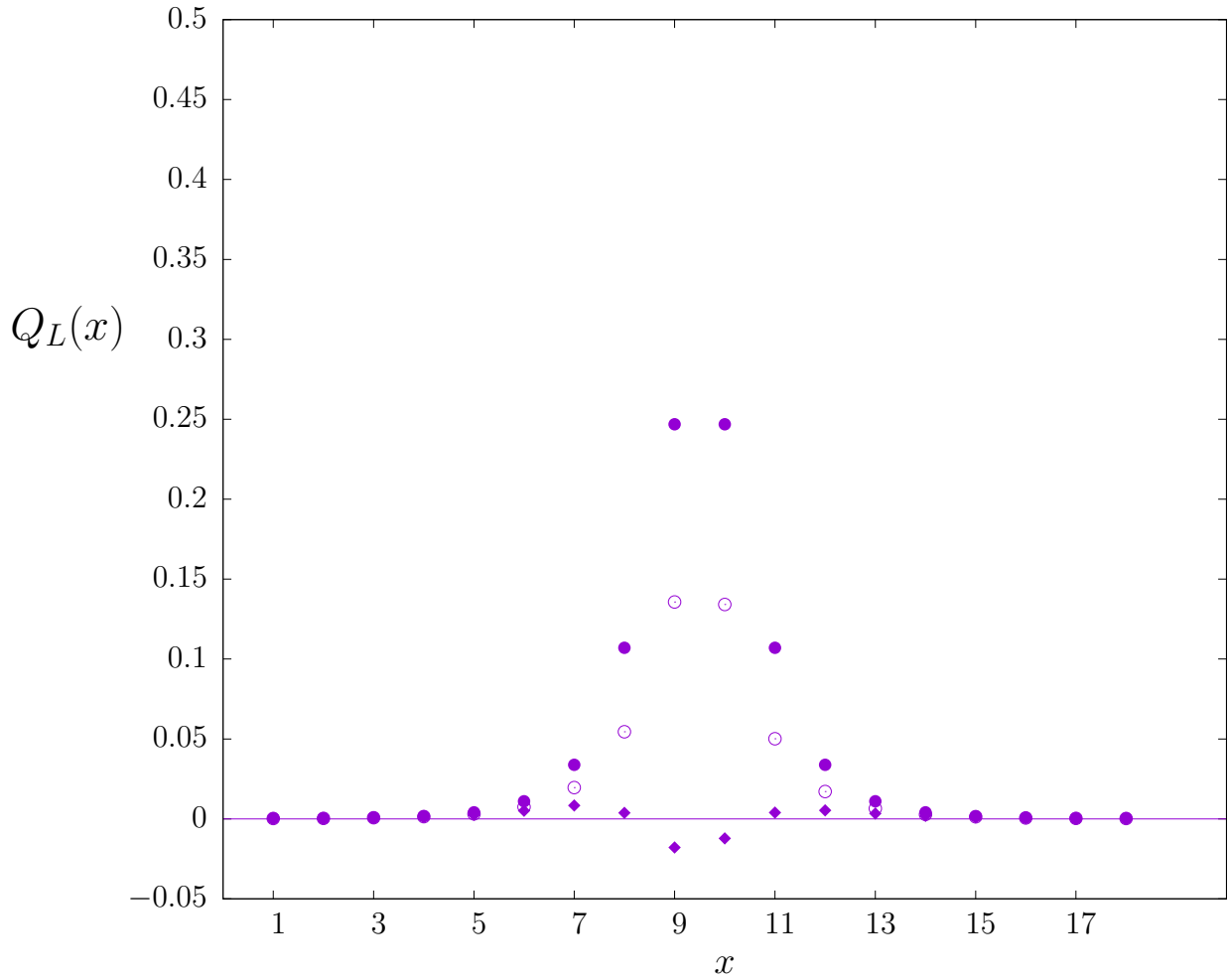}
\end	{center}
\caption{Cooling a classical instanton with $\rho=2a$ on an $18^4$ lattice.
  Profile in $x$ of the topological charge after no cools ($\bullet$), 16 cooling sweeps ($\circ$),
  20 cooling sweeps ($\blacklozenge$).}
\label{fig_plot_Iprofcool_r2l18}
\end{figure}

\begin{figure}[htb]
\begin	{center}
\leavevmode
\input	{plot_QLhistrun_sm8_l36b6.70.tex}
\end	{center}
\caption{Histogram of the values of the lattice topological charge, $Q_L$, obtained from
  a sequence of $SU(3)$ lattice fields that have been subjected to 7 smearings, $i_s=7$.
  The fields are grouped into ensembles labelled by the topological charge $Q$ that
  one infers after 20 cooling sweeps. Those corresponding to odd and even $Q$ values are
  labelled by $\bullet$ and $\circ$ respectively.  Zero values suppressed.
  These fields on $36^344$ lattices are generated at $\beta=6.70$.}
\label{fig_QLhistrun_sm8_l36b6.70}
\end{figure}

\begin{figure}[htb]
\begin	{center}
\leavevmode
\input	{plot_QLhistrun_sm8_l18b6.235.tex}
\end	{center}
\caption{Histogram of the values of the lattice topological charge, $Q_L$, obtained from
  a sequence of $SU(3)$ lattice fields that have been subjected to 7 smearings, $i_s=7$.
  The fields are grouped into ensembles labelled by the topological charge $Q$ that
  one infers after 20 cooling sweeps. Those corresponding to odd and even $Q$ values are
  labelled by $\bullet$ and $\circ$ respectively.  Zero values suppressed.
  These fields are on $18^326$ lattices generated at $\beta=6.235$.}
\label{fig_QLhistrun_sm8_l18b6.235}
\end{figure}

\begin{figure}[htb]
\begin	{center}
\leavevmode
\input	{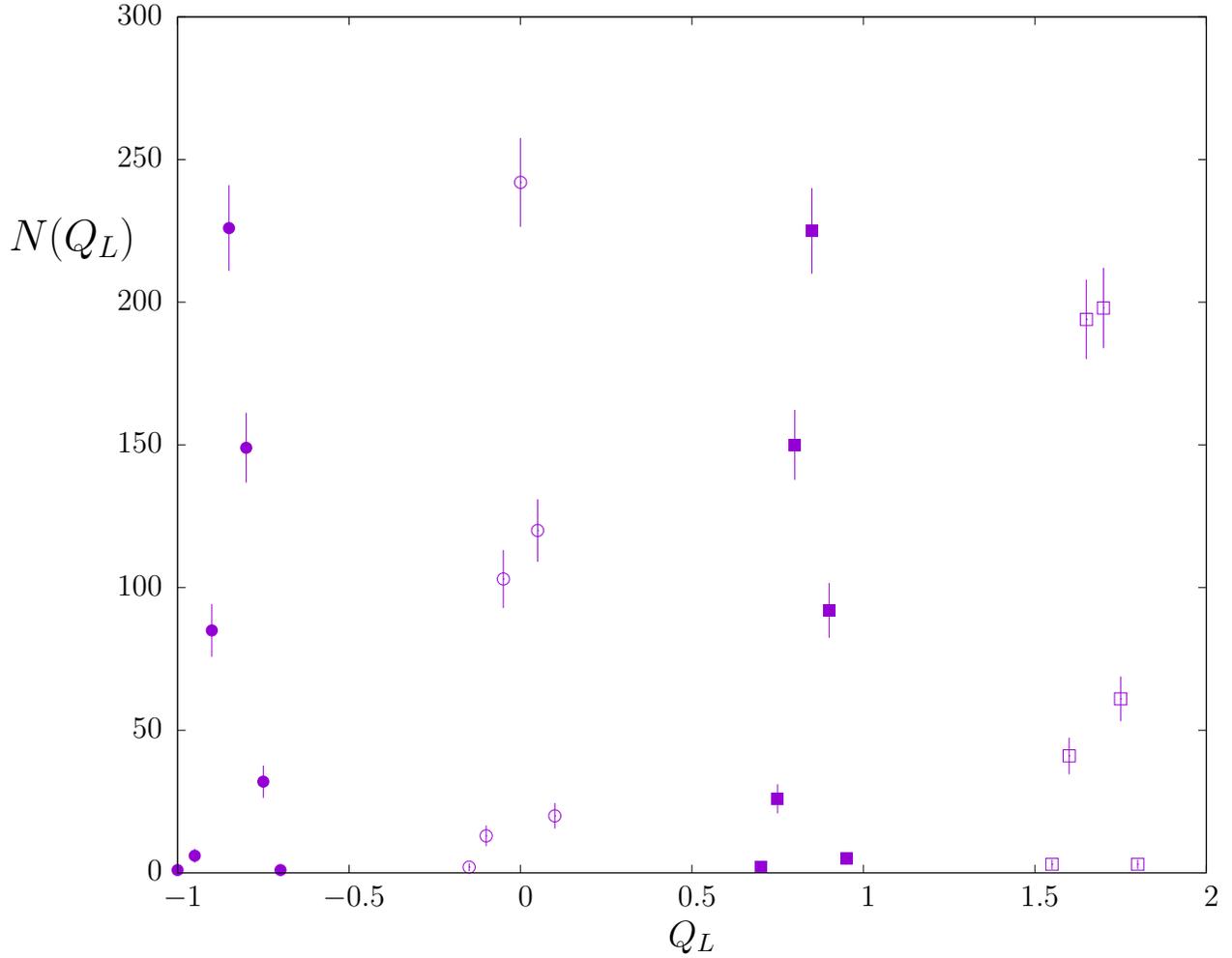}
\end	{center}
\caption{Histogram of the values of the lattice topological charge, $Q_L$, obtained from
  a sequence of $SU(8)$  lattice fields that have been subjected to 7 smearings, $i_s=7$.
  The fields are grouped into ensembles labelled by the topological charge $Q$ that
  one infers after 20 cooling sweeps: $Q=-1$ ($\bullet$), $Q=0$ ($\circ$), 
   $Q=+1$ ($\blacksquare$), $Q=+2$ ($\square$) respectively. Zero values suppressed.
  These fields are on $16^326$ lattices generated at $\beta=46.70$.}
\label{fig_QLhistrun_sm8_l16b46.70}
\end{figure}

\begin{figure}[htb]
\begin	{center}
\leavevmode
\input	{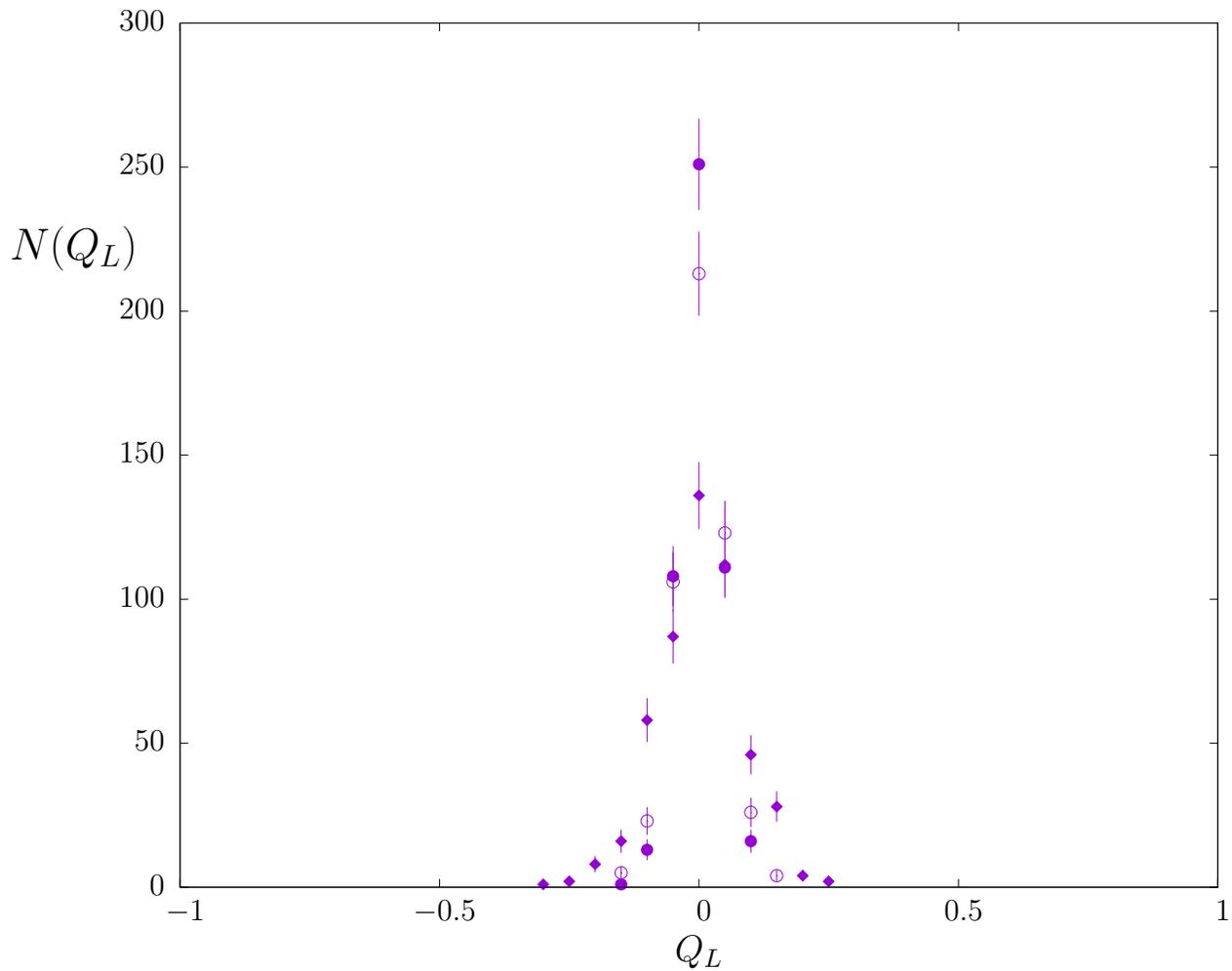}
\end	{center}
\caption{Histogram of the values of the lattice topological charge, $Q_L$, obtained from
  a sequence of 500 $SU(8)$  fields that have $Q=0$, with  $Q_L$ calculated after various
  numbers of smearing steps:
  4 ($\blacklozenge$), 6 ($\circ$),8 ($\bullet$).
  The fields are on $16^326$ lattices generated at $\beta=46.70$.}
\label{fig_QLhistrun_smQ0_l16b46.70}
\end{figure}

\clearpage

\begin{figure}[htb]
\begin	{center}
\leavevmode
\input	{plot_Q1proft_sm6vsr10Kl16b46.70.tex}
\end	{center}
\caption{Profiles in $t$ of the lattice topological charge density from a single $Q=+1$
  $SU(8)$ gauge field on an $16^324$ lattice, generated at $\beta=46.70$. Shown is the  profile 
  of the field after 5 smearing steps  ($\bullet$) compared to the average profile ($\circ$)
  obtained from an ensemble of 10000 fields each 3 heat bath sweeps from the
  given field, normalised to a common value of $Q_L$.} 
\label{fig_Q1proft_sm6vsr10Kl16b46.70}
\end{figure}

\begin{figure}[htb]
\begin	{center}
\leavevmode
\input	{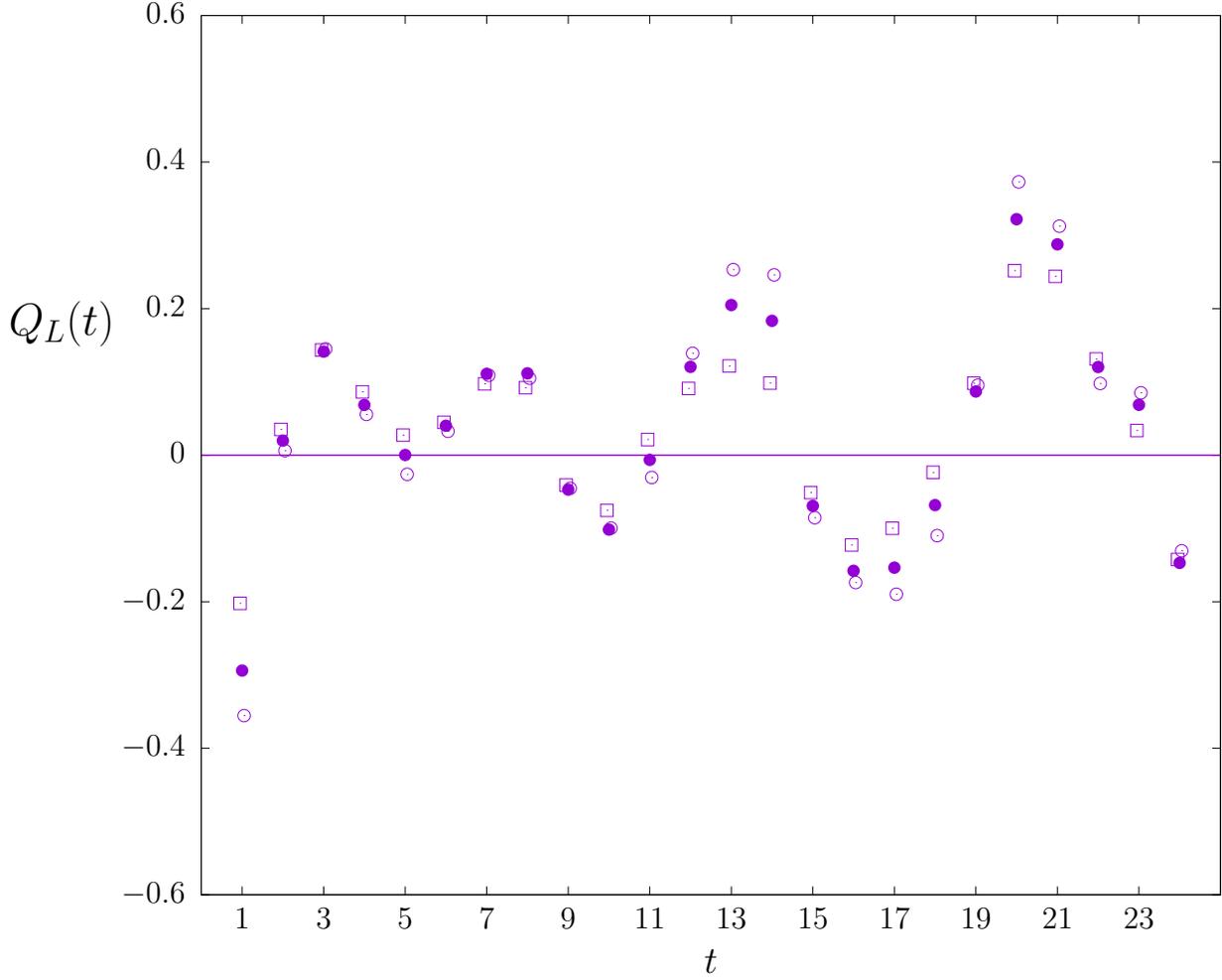}
\end	{center}
\caption{Profiles in $t$ of the lattice topological charge density from a single $Q=+1$
  $SU(8)$ gauge field on an $16^324$ lattice, generated at $\beta=46.70$. Shown is the  profile 
  of the field after 4($\circ$), 5($\bullet$) and 8($\square$) smearing steps.}
\label{fig_Q1proft_sm568l16b46.70}
\end{figure}

\begin{figure}[htb]
\begin	{center}
\leavevmode
\input	{plot_Q-1proft_sm6vsr10Kl26b6.50.tex}
\end	{center}
\caption{Profiles in $t$ of the lattice topological charge density from a single $Q=-1$
  $SU(3)$ gauge field on an $26^338$ lattice, generated at $\beta=6.50$. Shown is the  profile 
  of the field after 5 smearing steps  ($\bullet$) compared to the average profile ($\circ$)
  obtained from an ensemble of 10000 fields each 3 heat bath sweeps from the
  given field, normalised to a common value of $Q_L$.} 
\label{fig_Q-1proft_sm6vsr10Kl26b6.50}
\end{figure}




\clearpage




\begin{figure}[htb]
\begin	{center}
\leavevmode
\input	{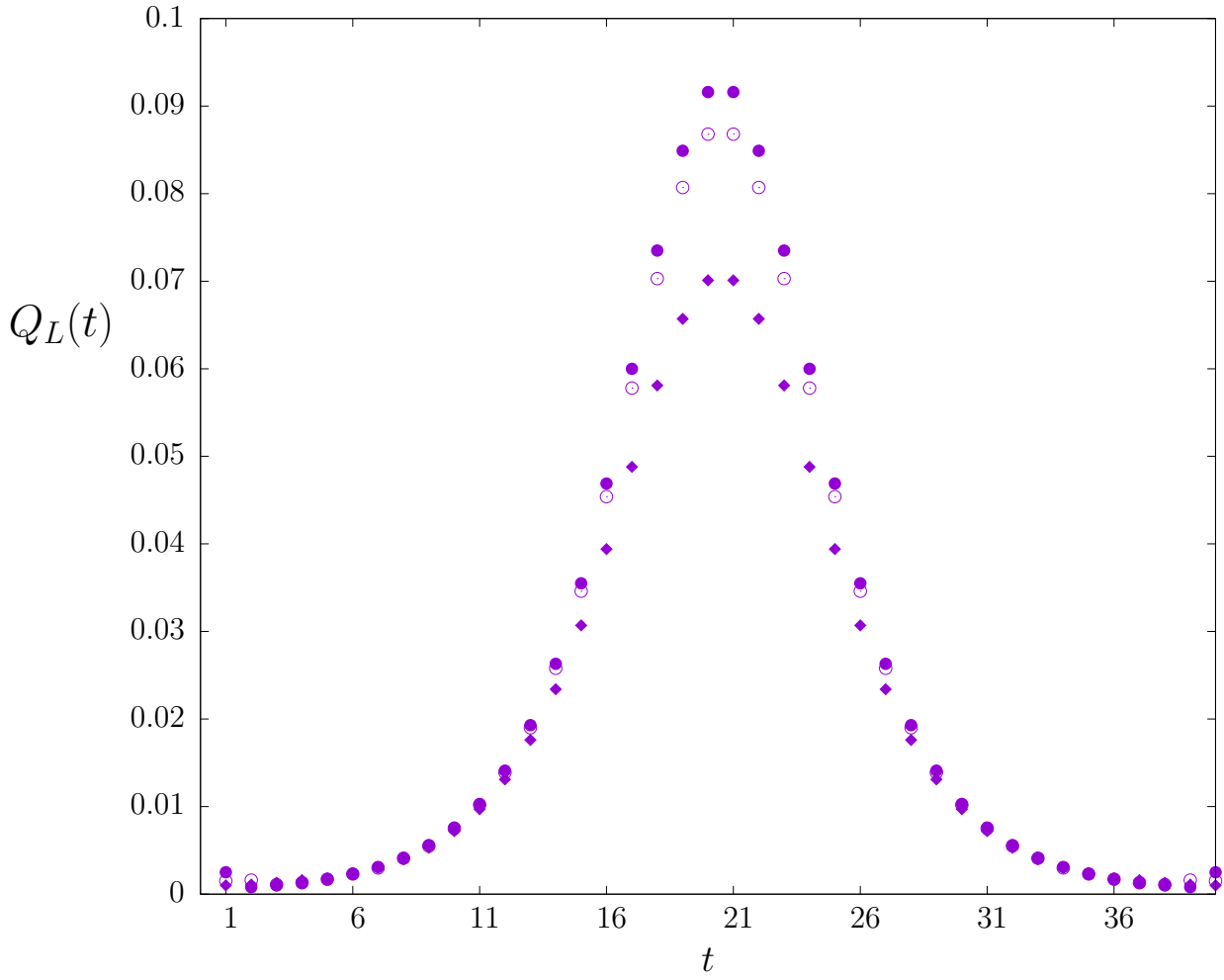}
\end	{center}
\caption{Profile of the integrated topological charge $Q_L(t)$ in each timeslice $t$ of a
  classical instanton with size $\rho=8a$ on a $40^4$ lattice. With no blocking, $\bullet$,
  single blocking, $\circ$, and double blocking, $\blacklozenge$.}
\label{fig_Iprofbl_r8l40}
\end{figure}

\begin{figure}[htb]
\begin	{center}
\leavevmode
\input	{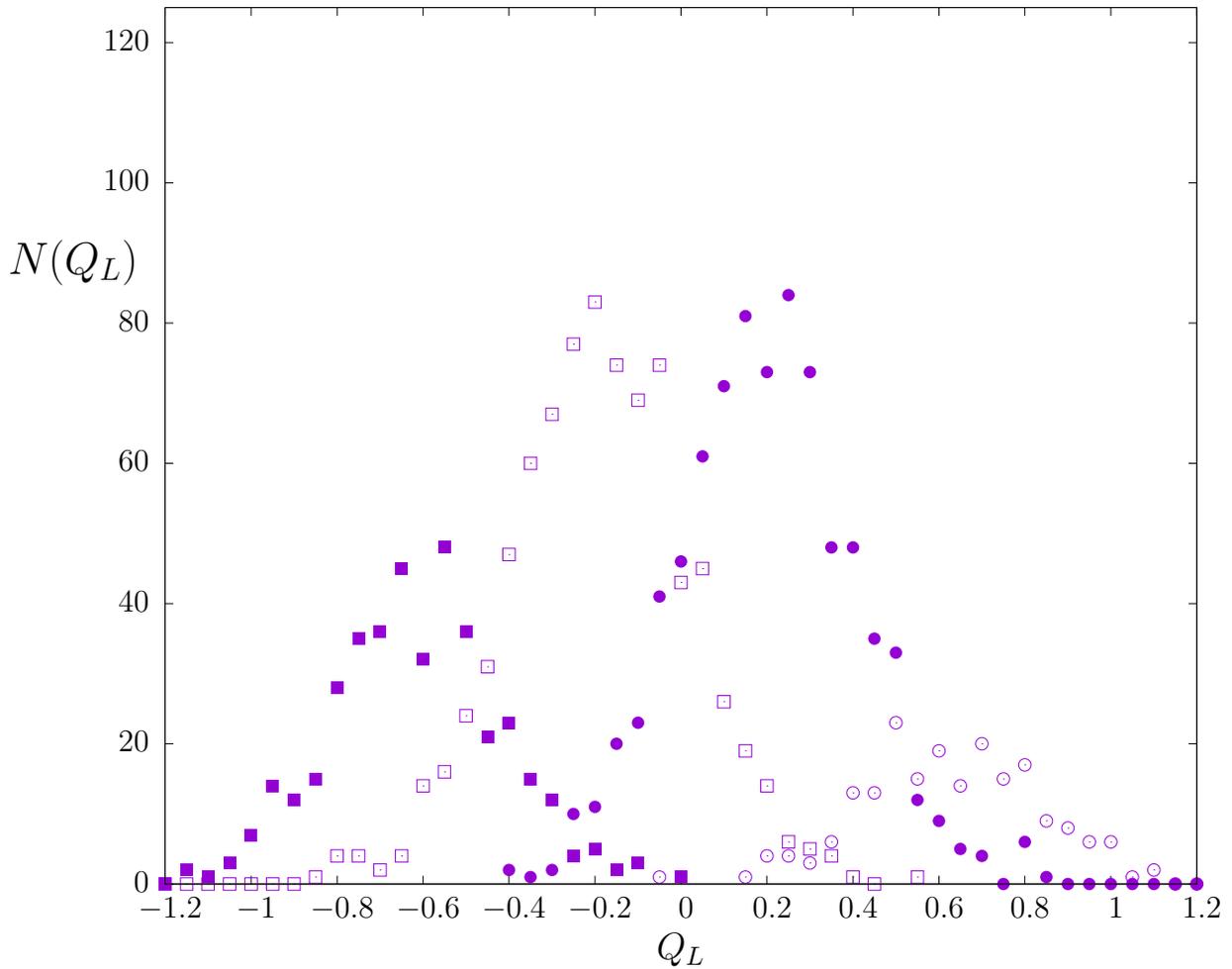}
\end	{center}
\caption{$Q_L$ on a sequence of doubly blocked $SU(3)$ gauge fields
  generated at $\beta=6.70$. Histograms of fields with $Q=3$ ($\circ$),
  $Q=1$ ($\bullet$), $Q=-1$ ($\square$),$Q=-3$ ($\blacksquare$), where
  the value of $Q$ is obtained after 20 cooling sweeps of the unblocked fields.}
\label{fig_Qhistbl3_b6.70b}
\end{figure}

%


%

\clearpage

\end{document}